\documentclass[10pt,journal]{IEEEtran}
\usepackage{cite}
\usepackage{amssymb,amsfonts}
\usepackage{algorithmic}
\usepackage{graphicx}
\usepackage[table,xcdraw]{xcolor}
\usepackage{textcomp}
\usepackage{fixltx2e}
\usepackage{soul}
\usepackage{xcolor}
\usepackage{colortbl}
\usepackage{enumerate}
\usepackage{siunitx}
\usepackage{tabularx}
\usepackage{booktabs}
\usepackage[hyphens]{url}
\usepackage{hyperref,color,soul,multirow}
\usepackage{epstopdf}
\usepackage{array}
\usepackage[cmex10]{amsmath}
\usepackage{float}
\usepackage[flushleft]{threeparttable} 
\usepackage{booktabs,caption}
\usepackage{comment}
\usepackage{epsfig,bm,graphicx,setspace}
\usepackage{algorithmic,algorithm}
\usepackage{subcaption}
\usepackage{textcomp}
\usepackage[utf8]{inputenc}
\usepackage{csquotes}
\usepackage{dirtytalk}
\usepackage{cite}
\usepackage{dblfloatfix} 
\usepackage{url}
\usepackage{longtable}
\usepackage{nicefrac}
\usepackage{tabu}
\usepackage{amsmath}
\usepackage{pifont}
\newcommand{\cmark}{\ding{51}}

\graphicspath{{../}}
\DeclareGraphicsExtensions{.eps,.pdf,.jpeg,.png}

\def\BibTeX{{\rm B\kern-.05em{\sc i\kern-.025em b}\kern-.08em
    T\kern-.1667em\lower.7ex\hbox{E}\kern-.125emX}}
\usepackage{etoolbox}

\newcolumntype{C}[1]{>{\centering\arraybackslash}m{#1}}
\newcolumntype{P}[1]{>{\centering\arraybackslash}p{#1}}

\begin{document}

\title{A Comprehensive Survey on the Applications of Blockchain for Securing Vehicular Networks}

\author{Tejasvi Alladi$^*$,~\IEEEmembership{Senior Member,~IEEE}, Vinay Chamola$^*$,~\IEEEmembership{Senior Member,~IEEE}, Nishad Sahu, Vishnu Venkatesh, Adit Goyal and Mohsen Guizani,~\IEEEmembership{Fellow~IEEE}
\thanks{$^*$ T. Alladi and V. Chamola have equal contribution}
\thanks{T. Alladi is with the Department of Computer Science and Information Systems, BITS-Pilani, Pilani Campus, 333031, India. (e-mail: tejasvi.alladi@pilani.bits-pilani.ac.in).}
\thanks{N. Sahu, V. Venkatesh, and V. Chamola are with the Department of Electrical and Electronics Engineering \& APPCAIR, BITS-Pilani, Pilani Campus, 333031, India. (e-mail: \{h20160215, f20171154, vinay.chamola\}@pilani.bits-pilani.ac.in).}
\thanks{Adit Goyal is with the Department of Computer Science and IT, Jaypee Institute of Information Technology, Noida, India 201304 (email: aditgoyal@hotmail.com)}
\thanks{Mohsen Guizani is with the Department of Computer Science, Qatar University, Qatar (e-mail: mguizani@ieee.org).}
}
\maketitle

\begin{abstract}
\textcolor{black}{Vehicular networks promise features such as traffic management, route scheduling, data exchange, entertainment, and much more. With any large-scale technological integration comes the challenge of providing security. Blockchain technology has been a popular choice of many studies for making the vehicular network more secure. Its characteristics meet some of the essential security requirements such as decentralization, transparency, tamper-proof nature, and public audit. This study catalogues some of the notable efforts in this direction over the last few years. We analyze around 75 blockchain-based security schemes for vehicular networks from an application, security, and blockchain perspective. The application perspective focuses on various applications which use secure blockchain-based vehicular networks such as transportation, parking, data sharing/ trading, and resource sharing. The security perspective focuses on security requirements and attacks. The blockchain perspective focuses on blockchain platforms, blockchain types, and consensus mechanisms used in blockchain implementation. We also compile the popular simulation tools used for simulating blockchain and for simulating vehicular networks. Additionally, to give the readers a broader perspective of the research area, we discuss the role of various state-of-the-art emerging technologies in blockchain-based vehicular networks. Lastly, we summarize the survey by listing out some common challenges and the future research directions in this field.}
\end{abstract}

\vspace{-0.1in}

\section{Introduction}
\label{Intro}

\begin{figure*}[]
        \centerline{\includegraphics[width = 2\columnwidth]{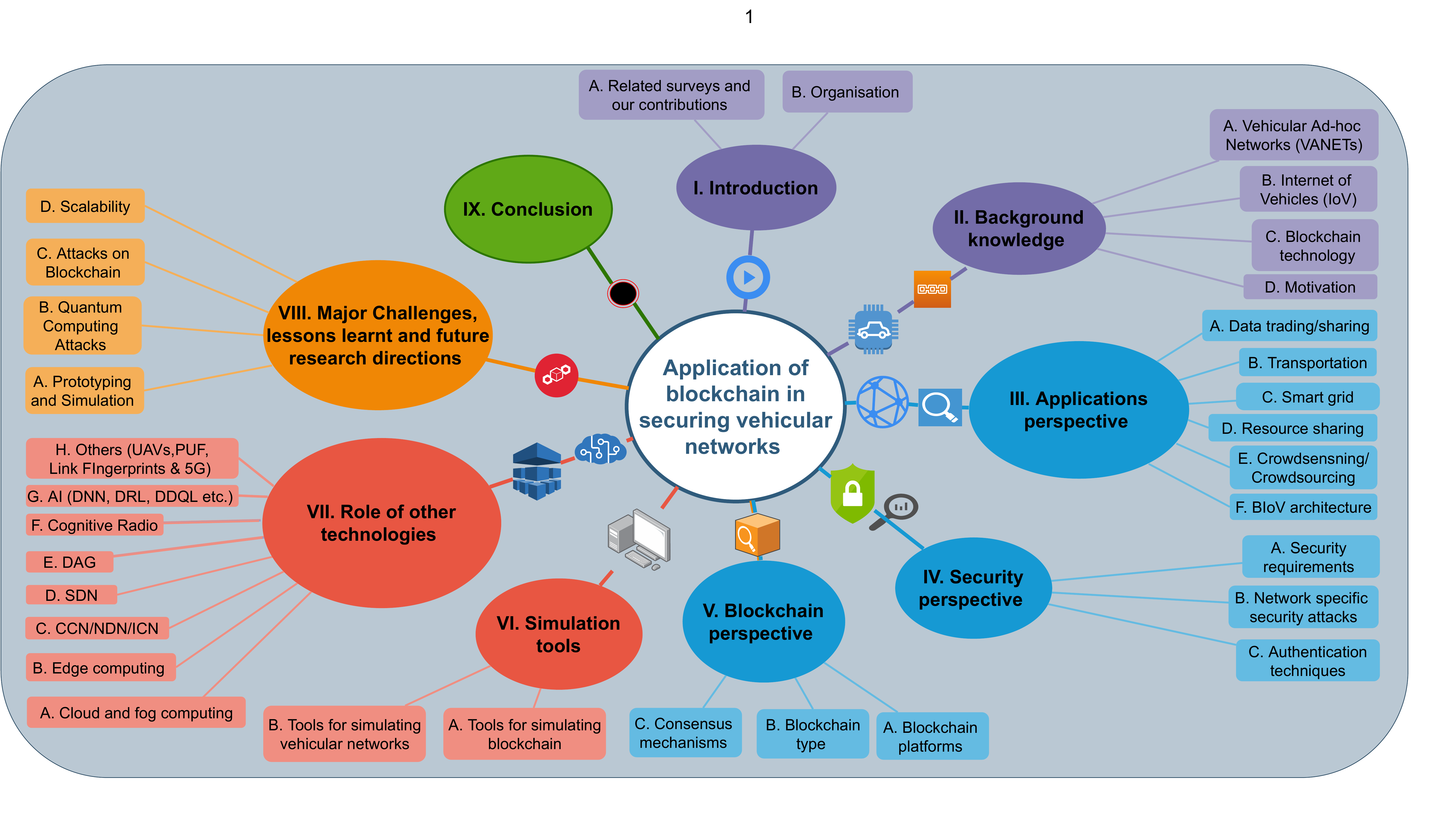}}
        \caption{\textcolor{black}{Overview of this survey.}}
        \label{fig:survey_overview}
\end{figure*}

The number of vehicles moving on the roads reached $1$ billion in $2010$. Experts predict that this could reach $2 - 2.5$ billion by $2050$ and that a large part of it will form futuristic vehicular network traffic comprising connected vehicles. Modern vehicles are no longer mere thermo-mechanical machines but are a combination of sophisticated hardware and software \cite{sahu2012bahg}. They have GPS facilities, wireless communication devices, entertainment systems, advanced sensing mechanisms, visual aids, automatic alarm systems, and many more features, thus involving a lot of data processing and connectivity. Since travelling on roads is seldom a solitary endeavour, the next logical step in technological evolution is to enable individual vehicles to communicate and coordinate with each other. A connected vehicular network promises features such as traffic management, route scheduling, data exchange, entertainment, and much more. Vehicular Ad-hoc Network (VANET) is a way of achieving this. A VANET is a type of Mobile Ad-hoc Network (MANET) designed for network nodes that are constantly in motion supported by road-side infrastructure called Road Side Units (RSUs) that theoretically can be used to serve a large variety of features, anywhere from urgent collision notifications to interlinking geographically separate VANETs \cite{pathan2016security, elsadig2016vanets}. Vehicles in a VANET can communicate with each other via Vehicle-to-Vehicle (V2V) communication and with RSUs via Vehicle-to-Infrastructure (V2I) communication. Researchers believe that with the help of VANET technology we can overcome a lot of issues including, but not limited to, crash prevention and safety \cite{mehra2020reviewnet}, driver assistance, and freeway management. For example, in case of an accident, the vehicles may broadcast the information to distant vehicles that may be planning to use the same route. Road safety is also a growing concern. The number of road traffic deaths across the globe was reported to be around $1.25$ million by the Global Health Observatory (GHO) data \cite{who_road_safety}. VANETs, believed to be an effective road safety solution also have the potential to contribute to efficient traffic management in urban scenarios. Also, with the adoption of IoT in VANETs, a new paradigm called the Internet of Vehicles (IoV) is being discussed as a futuristic vehicular network. However, there is also a downside to the growing number of V2V and V2I communication links as part of the VANET and the IoV paradigm, in terms of network security \cite{alladi2020consumer,alladi2020industrial}. In this regard, several research articles have been proposed addressing security issues in vehicular networks \cite{alladi2020lightweight}.The number of vehicles moving on the roads reached $1$ billion in $2010$. Experts predict that this could reach $2 - 2.5$ billion by $2050$ and that a large part of it will form futuristic vehicular network traffic comprising connected vehicles. Modern vehicles are no longer mere thermo-mechanical machines but are a combination of sophisticated hardware and software \cite{sahu2012bahg}. They have GPS facilities, wireless communication devices, entertainment systems, advanced sensing mechanisms, visual aids, automatic alarm systems, and many more features, thus involving a lot of data processing and connectivity. Since travelling on roads is seldom a solitary endeavour, the next logical step in technological evolution is to enable individual vehicles to communicate and coordinate with each other. A connected vehicular network promises features such as traffic management, route scheduling, data exchange, entertainment, and much more. Vehicular Ad-hoc Network (VANET) is a way of achieving this. A VANET is a type of Mobile Ad-hoc Network (MANET) designed for network nodes that are constantly in a motion supported by road-side infrastructure called Road Side Units (RSUs) that theoretically can be used to serve a large variety of features, anywhere from urgent collision notifications to interlinking geographically separate VANETs \cite{pathan2016security, elsadig2016vanets}. Vehicles in a VANET can communicate with each other via Vehicle-to-Vehicle (V2V) communication and with RSUs via Vehicle-to-Infrastructure (V2I) communication. Researchers believe that with the help of VANET technology, we can overcome a lot of issues including, but not limited to, crash prevention and safety \cite{mehra2020reviewnet}, driver assistance, and freeway management. For example, in case of an accident, the vehicles may broadcast the information to distant vehicles that may be planning to use the same route. Road safety is also a growing concern. The number of road traffic deaths across the globe was reported to be around $1.25$ million by the Global Health Observatory (GHO) data \cite{who_road_safety}. VANETs believed to be an effective road safety solution, also have the potential to contribute to efficient traffic management in urban scenarios. Also, with the adoption of IoT in VANETs, a new paradigm called the Internet of Vehicles (IoV) is being discussed as \textcolor{black}{a} futuristic vehicular network. However, there is also a downside to the growing number of V2V and V2I communication links as part of the VANET and the IoV paradigm, in terms of network security \cite{alladi2020consumer, alladi2020industrial}. In this regard, several research articles have been proposed addressing security issues in vehicular networks \cite{alladi2020lightweight}.

\subsection{Related Surveys}
We first discuss the existing surveys on blockchain in vehicular networks and contrast them with the present survey. Baldini et al. presented a short review of the applications of distributed ledger technologies for the evolution of road transportation \cite{baldini2020review}. Another work by authors in \cite{bao2020survey} surveyed the applications of blockchain in the energy sector, with one application area being electric vehicles. A broad survey by Xie et al. \cite{xie2019survey} on applications of blockchain in smart cities discusses several aspects of smart cities, with smart transportation being one of the major areas of smart cities considered. Mollah et al. \cite{mollah2020blockchain} recently proposed a comprehensive survey for blockchain applications in the IoV network. The authors in \cite{butt2019privacy} presented a review of blockchain-based privacy management techniques in social IoV networks. Many of these surveys focus more on the application areas and the frameworks and mention the security features that are implicit with a blockchain-based framework. However, a study that analyses security from a more rigorous requirement-oriented viewpoint that discusses how precisely blockchain meets those requirements would be timely. Other perspectives, like applications and an overview of how blockchain platforms have been used, are provided to make this study self-contained for a reader. In summary, this paper aims to help the readers get an overview of the security challenges and requirements in vehicular networks and the blockchain techniques developed to mitigate these challenges. A list of major survey and review works done on blockchain applications in vehicular networks in the last few years (as discussed above) is given in Table \ref{table_relatedworks}.
 
\begin{table*}[!t]
\caption{Related surveys on blockchain applications in vehicular networks}
\centering
\resizebox{1\textwidth}{!}{
 \begin{tabular}{|l|l|l|}
 					\hline \rowcolor[gray]{0.7}
 					\textbf{Year} & \textbf{Author} & \textbf{Contributions} \\
					
 					\hline
 	2019 & Xie et al. \cite{xie2019survey}
  & Surveys the application of blockchain in smart city scenarios \\
 				    \hline
    2019 &  Butt et al., \cite{butt2019privacy}
  & Presents a review of blockchain based solutions for managing privacy in social IoV  \\
                 	\hline
 		2020 & Baldini et al. \cite{baldini2020review}
  &  Reviews the use of distributed ledger in road transport evolution \\
 		          \hline
 	2020 & Boa et al. \cite{bao2020survey}
 &    Surveys the application of blockchain in the energy sector \\
 		            \hline
     					2020 & Mollah et al. \cite{mollah2020blockchain}
  & Surveys usage of blockchain in IoV and Intelligent transportation systems (ITS)      \\
 				    \hline
 	2021 & This Survey  & Surveys the use of blockchain with a focus on security and privacy in VANETs \\ \hline
 \end{tabular}	}
\label{table_relatedworks}
\end{table*}

\subsection{Motivation}
Vehicles in the network share and access data with other vehicles (V2V), humans (V2H), sensors (V2S), infrastructure (V2I), or anything else (V2X) to provide a variety of services such as traffic management, energy exchange, accident avoidance, and improved driving experience. There are two aspects to blockchain technology that makes it particularly useful for vehicular network applications. One is the structural aspect of the blockchain - the technology itself is designed to provide security services and data integrity in a way that does not depend on the trustworthiness of a third party. The other aspect is the functionality of smart contracts, which provide a mechanism to carry out complex tasks and allow a large number of nodes or vehicles to interact intelligently.

Data sharing is a key feature of the IoV - at all times, there is a large amount of data that is being generated and distributed across the network. Vehicles access that data to make decisions or upload it to access cloud services. Along with the data comes the associated problems of security. The issue of security will be addressed in-depth in the coming sections of this paper. Smart contracts have expanded the functionality of blockchain beyond simple distributed ledger mechanisms. They can facilitate energy sharing between electric vehicles when blockchain is used for smart grid applications \cite{su2018secure, zhou2019blockchain}, optimizing demand and supply for data trading (or computing resource sharing) for IoV applications \cite{song2020blockchain,liu2019novel,chen2019secure,chai2019proof}, and task scheduling \cite{fan2018research}. They can also be used for security purposes, when only a certain number of nodes need to be given the authorization to transact \cite{song2020blockchain, wang2019improved}. Authentication plays an important role in a vehicular network, due to the underlying dynamic topology with vehicles joining and leaving frequently. Several research works \cite{lu2019blockchain,zheng2019traceable,hu2019blockchain,wang2019improved} provide frameworks for authentication in VANETs. Traffic control mechanisms are also an area where the IoV can use blockchains effectively. Cheng et al. give a model for access control of data using attribute-based blockchains, where only vehicles conforming to certain attributes (road, the direction of travel, etc) can access certain information \cite{cheng2019sctsc}. Platooning is a model where different vehicles form groups for navigation and travel in those groups to ease traffic congestion, and provide more easily manageable coordination between vehicles \cite{chen2019smart}.

In this survey, each research work was analyzed based on the problem it addressed. From that analysis emerged some broad research thrust areas that the research community is currently working on. Based on this trend, future research in this field shall be closely related to one or more of these thrust areas. Table \ref{tab:research_problems} illustrates the major research areas with some works containing the problem addressed in that research direction and the proposed solution.

The main contributions of this work are as follows:
\begin{enumerate}[i.]
 \item {The study analyses recent research in the area of blockchain-based vehicular networks from the application perspective, where the studies are classified based on the application area considered.}
 \item {The study also analyses the research from a security perspective wherein studies are classified based on security requirements met, security attacks protected against, authentication techniques used, and security proofs showed.}
 \item {Recent advances are also examined from a blockchain perspective, categorised by blockchain platforms and consensus.}
 \item {The study discusses various simulation tools which have been used in blockchain-based vehicular network studies.}
 \item {We also provide some insight on the role of other state-of-the-art technologies including, but not limited to, cloud computing, fog computing, edge computing, Software Defined Networking (SDN), Named Data Networking (NDN), Artificial Intelligence, 5G, etc. in securing blockchain-based vehicular networks.}
 \item {Based on the survey we present some major challenges and probable future research directions in this field.}
\end{enumerate}

\subsection{Organisation}
The rest of the paper is organised as follows. In Section \ref{sec:bckg}, we present the background knowledge on VANETs, IoV networks, and blockchain technology. We discuss the blockchain-based security frameworks for vehicular networks from an applications perspective in Section \ref{sec:nw_app}. Section \ref{sec:security_perspective} discusses the categorisation of the different blockchain-based works from the security perspective. These works are further categorised from the blockchain perspective in section \ref{sec:bc_perspective}. A compilation of different simulation tools used is presented in Section \ref{sec:simulation}. We discuss the role of other state-of-the-art technologies in securing blockchain-based vehicular frameworks in Section \ref{sec:role_of_other_tech}. Section \ref{sec:future} describes existing challenges/open issues in using blockchain to secure vehicular networks and presents some future research directions. Finally, we conclude the paper in Section \ref{sec:conc}. The organisational overview of this survey is also shown in Fig. \ref{fig:survey_overview}.

\begin{table}[]
\caption{Major abbreviations used in the survey.}
\label{tab:major_abbr}
\begin{tabular}{|l |l|}
\hline
\cellcolor[HTML]{C0C0C0}\textbf{Notation} & \cellcolor[HTML]{C0C0C0}\textbf{Meaning} \\ \hline
{\color[HTML]{000000} 3GPP} & 3rd Generation Partnership Project \\ \hline
{\color[HTML]{000000} BFT} & Byzantine Fault Tolerance \\ \hline
{\color[HTML]{000000} BIoV} & Blockchain-based Internet of Vehicles \\ \hline
{\color[HTML]{000000} CA} & Certificate Authority \\ \hline
{\color[HTML]{000000} CCN} & Content Centric Networking \\ \hline
{\color[HTML]{000000} DAG} & Directed Acrylic Graph \\ \hline
{\color[HTML]{000000} DSRC} & Dedicated Short-Range Communications \\ \hline
{\color[HTML]{000000} GPR} & Gaussian Process Regression \\ \hline
{\color[HTML]{000000} IoEV} & Internet of Electric Vehicles \\ \hline
{\color[HTML]{000000} ITS} & Intelligent Transportation Systems \\ \hline
{\color[HTML]{000000} PBFT} & Practical Byzantine Fault Tolerance \\ \hline
{\color[HTML]{000000} PKI} & Public Key Infrastructure\\ \hline
{\color[HTML]{000000} PoW} & Proof of Work \\ \hline
{\color[HTML]{000000} SDN} & Software Defined Networking \\ \hline
{\color[HTML]{000000} SoC} & State of Charge ( of EVs) \\ \hline
{\color[HTML]{000000} TM} & Trace Managers \\ \hline
{\color[HTML]{000000} V2CH} & Vehicle to Cluster Head \\ \hline
{\color[HTML]{000000} V2I} & Vehicle to Infrastructure \\ \hline
{\color[HTML]{000000} V2V} & Vehicle to Vehicle \\ \hline
{\color[HTML]{000000} V2X} & Vehicle to Everything \\ \hline
{\color[HTML]{000000} VANET} & Vehicular Ad-hoc Networks \\ \hline
{\color[HTML]{000000} VCC} & Vehicular Cloud Computing \\ \hline
{\color[HTML]{000000} VEC} & Vehicular Edge Computing \\ \hline
{\color[HTML]{000000} VFS} & Vehicular Fog Services \\ \hline
{\color[HTML]{000000} WAVE} & Wireless Access in Vehicular Environment \\ \hline
{\color[HTML]{000000} ZKP} & Zero Knowledge Proof \\ \hline
\end{tabular}
\end{table}

\section{Background Knowledge}
\label{sec:bckg}
This section gives a brief background on VANETs, Internet of Vehicles, and blockchain technology.

\subsection{Vehicular Ad-Hoc Networks}
The idea behind a vehicular network is to take wireless technology that is used to network computers and adapt it to vehicles. The technical term for vehicular networks is Vehicular Ad-Hoc Networks (VANETs). Ad-hoc networks do not have a defined infrastructure, thus the network topology must be decided by the nodes in the network through cooperative mechanisms \cite{zanjireh2015survey}. In other words, there is no central authority and the nodes themselves behave as routers and are responsible for propagating information in the network.

\subsection{Internet of Vehicles}
The Internet of Vehicles can be seen as an extension of VANET, where the ability to network vehicles is used to connect to the internet or to cloud services and create Intelligent Transportation Systems (ITS) based on an IP-connected infrastructure. VANETs provide data connectivity; and the IoV includes the processing of that data on a very large scale, to provide essential services. The ideal IoV network would be a seamless integration of vehicles, the environment, and humans on the roads which would increase transport safety and efficiency on a city-wide or nationwide scale. Big data, cloud computing, and artificial intelligence are technologies that could be used in conjunction with VANETs to achieve this goal. An abstract architecture of an IoV network was first proposed by Yang et al \cite{yang2014overview}.

The IoV applications can be divided into two broad categories, namely, safety and commercial applications. Safety applications are essential services such as collision avoidance, speed limit information, and emergency braking information among others. These applications require low transmission latency in the network since they are often time-sensitive. Commercial applications include providing weather and traffic information, streaming of media, instant messaging, or any other service that enhances the driving experience. The important distinction is that safety applications are essential and are hard real-time applications; they must not be disturbed by commercial applications.

Some examples of IoV applications are given as follows: Olaverri-Monreal et al. propose a video-streaming technology that improves the visibility of the driver visibility and also supports overtaking in challenging scenarios \cite{olaverri2010see}, Lin et al. propose an on-board diagnostic system based on GPS and 3G \cite{lin2009study}, and MobEyes proposed by Lee et al. is a smart system that exploits wireless-enabled vehicles to perform event sensing, by having the mobile nodes constantly generate data summaries by extracting features from their environment and sharing with neighbour nodes \cite{lee2006mobeyes}.

\subsection{Blockchain Technology}
\color{blue}
Blockchain is a new technology that is rapidly gaining traction in fields such as finance \cite{tapscott2017blockchain}, UAVs \cite{BC_UAV, fernandez2018uav,kapitonov2017blockchain,lei2019securing}, IoT \cite{dorri2017towards, novo2018blockchain,panarello2018blockchain, alladi2019blockchain }, smart cities \cite{biswas2016securing,sun2016blockchain}, smart grids \cite{alladi2019blockchain}, supply chain management \cite{tribis2018supply}, VANETs \cite{leiding2016self,lu2018bars}, and many others. It was originally proposed by S. Nakamoto in his whitepaper \cite{nakamoto2019bitcoin} on Bitcoin. Blockchain is a type of data structure holding records of digital transactions, formally known as a distributed ledger. Identical copies of the database exist across multiple different computing machines, called nodes in blockchain terminology, connected in a peer-to-peer network. Transactions being the fundamental units of blockchain, a definite number of transactions are stored in a block, and blocks are continuously appended in sequence to form a chain. It emphasizes the importance of decentralization where the majority of entities participating in the blockchain are assumed to be genuine and take the decision collectively with the help of the process known as a consensus mechanism.

Some of the core ideas on which blockchain is built are outlined below:
\begin{enumerate}
    \item \textbf{Digital Signatures:}
    Public key cryptography is one of the core concepts of blockchain technology. Each agent is assigned a private key and a public key. Anything encrypted using the private key can only be decrypted using the public key, and vice versa. The public key serves as an address for each node, and each digital asset is associated with its owner's public key \cite{pilkington2016blockchain}. The piece of data that needs to be transferred is signed using the private key. This can be used to authenticate information; if a piece of data is signed cryptographically using a private key, then the only thing that can decrypt it is the same user's public key. Blockchains commonly use elliptic curve digital signature algorithms \cite{johnson2001elliptic}.
    
    \item \textbf{Hashing:}
    Hashing algorithms are arguably the backbone of blockchain technology. The hash function is a type of cryptographic algorithm which takes an input of variable size and returns an output of fixed length, called a hash. SHA family (SHA-1 and SHA-2) are popular hashing algorithms. There are two conditions a good hash algorithm must obey:
    \begin{enumerate}
        \item It must be non-invertible; i.e., it should not be possible to retrieve the input given the output.
        \item The chances of two different inputs giving the same output hash must be very small.
    \end{enumerate}
    The reason this is useful for security is that a small input change will completely change the hash value, and that makes tampering evident.

    \item \textbf{Blocks:}
    Blocks are the constituent elements of blockchains, and they typically consist of a block body and block header. The block body contains transactions and a transaction counter. The block header contains different pieces of information, such as the Merkle tree root, the timestamp, block version, and the previous block's header's hash. These stored hash values provide immutability to the transactions. If a transaction in any block is changed, then it would change the block header, and the hash value will be different from the hash value stored in the successive block, and thus tampering is evident.
    
    Each block is validated through a consensus algorithm and added using a process that is necessarily expensive or difficult to perform but easy to validate - the immutability comes from the belief that malicious entities will not be able to meet the conditions for this hard-to-perform-but-easy-to-validate mining process, and therefore, cannot simply change the hash values of the blocks to cover up any tampering. The mining process needs to be performed for all subsequent blocks if a certain block is modified after creation and added to the chain, which is practically impossible. The blockchain is public, so participating nodes will be able to view but not modify the contents. A string of blocks appended in sequence form the blockchain.
    
\begin{figure}[h]
    \centerline{\includegraphics[width=9cm]{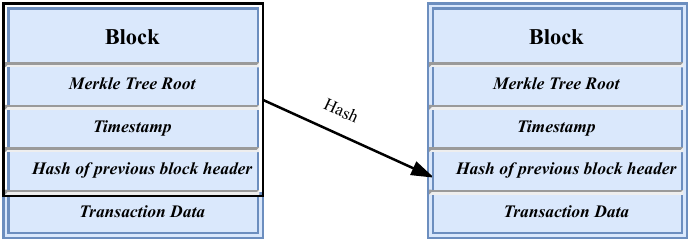}}
    \caption{Blocks in a blockchain}
    \label{fig:appsc}
    %\vspace{-0.2in}
\end{figure}

    \item \textbf{Consensus Algorithm: } 
    Nodes in the peer-to-peer network take the responsibility of verifying the transactions and adding them to the blockchain. This process is known as mining and is one of the most important elements of the blockchain network because it is responsible for its decentralized nature. The fundamental idea behind consensus is that nodes must undergo a process that is hard to perform yet easy to validate - discouraging malicious entities from acquiring the necessary conditions required to validate invalid transactions. 
\end{enumerate}

Putting it all together - suppose Alice desires to send a digital asset to Bob. Then Alice would have to sign the asset using her private key and broadcast a transaction request with the item and Bob's address. A miner, upon receiving the transaction, would bundle that transaction along with several other transactions in the block body. The miner would also create the block header and subsequently, broadcast the header to other blockchain nodes. These blockchain nodes then perform a pre-decided consensus algorithm. If the block is approved, then it is added as the latest block and all the nodes update the ledger to reflect the change. The fundamental role of the miner in all this is to collect, verify, and package transactions into a block, though the specifics of how they would do are dependent on the type of blockchain and consensus mechanisms agreed upon.

\begin{figure*}
    \centerline{\includegraphics[width =1.5 \columnwidth]{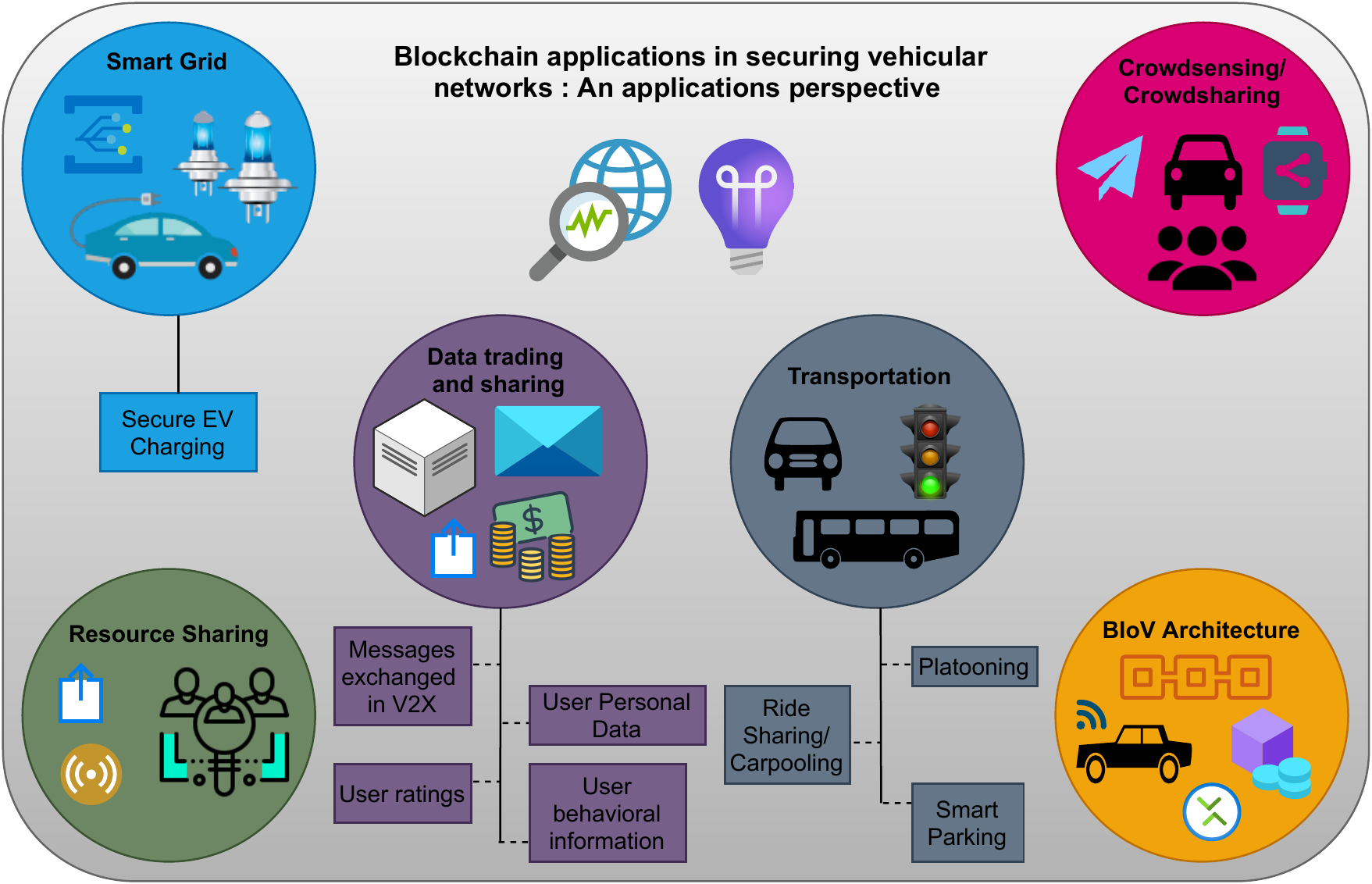}}
    \caption{Overview of the applications perspective section.}
    \label{fig:application_perspective}
\end{figure*}

There are two major categories of blockchains - permissioned and permissionless:
\begin{enumerate}[i.]
    \item \textbf{Permissionless blockchains} are public and open access; anyone is capable of joining the blockchain and take part in the consensus mechanism. Interested users having an internet connection can join become a part of the network, and participants’ identities are hidden which is a security concern.
    
    \item \textbf{Permissioned blockchains} place restrictions on the member nodes in terms of read access or participation in the consensus process, or both. This often helps in computation and network communication overhead, which is a major cause for delay in permissionless networks. 
\end{enumerate}

Smart contracts are pieces of computer code that can run on a blockchain to facilitate and enforce the terms of an agreement. First proposed by Nick Szabo \cite{nickszabosc} in 1997, the concept behind a smart contract is to execute the functions/ tasks of an agreement automatically when the specified conditions of a contract/ agreement are satisfied. 

\color{black}
\section{Categorisation based on Application Scenarios}
\label{sec:nw_app}
In this section, we categorise the different blockchain-based security works surveyed from the application perspective, i.e., based on the application area considered.

\subsection{Data Trading and Sharing}
The concept of data trading/ sharing is to treat data as a commodity, with vehicles being able to `purchase' and `sell' data from the network. From a very broad point of view, this is the fundamental concept behind all other blockchain-based IoV frameworks; whether it is computational information for resource sharing or battery level information for smart grid applications. Data must be shared for vehicles to coordinate with each other. However, we address it as a separate thrust area to delineate the research that focuses on optimizing data trading/sharing frameworks from research works that take data sharing for granted and focus on other aspects of the network. We classify the data in vehicular networks into four broad categories:

\begin{itemize}
    \item\textbf{Messages exchanged:} Vehicles communicate with each other by exchanging messages such as sensor data and traffic-related information.
    \item \textbf{User personal data:} User personal data refers to the user identity, name, e-wallet details, images, and videos - anything that can be used to identify a vehicle. It can also include important parameters of Electric Vehicles (EVs) such as State of Charge (SoC) of battery, battery capacity, and travel schedules, all of which decide the EV type.
    \item \textbf{User \textcolor{black}{behavioural} information:} This includes information about the trading preferences of the individuals when participating in a trading network (data trading, resource trading, and energy trading among others), individual's likes and dislikes - any information that can be used to predict how a user will behave.
    \item \textbf{User ratings:} Various frameworks assign trust ratings to the vehicles based on their previous history which can be used by other nodes to evaluate the user's trustworthiness. These ratings need to be secured against false ratings uploaded by malicious vehicles. Such data is shared when a vehicle leaves one vehicular network and joins another network.
\end{itemize}

\begin{table*}[]
\scriptsize
\caption{Research problems and proposed solutions \textcolor{blue}{of blockchain-based security works} for some common application areas.}
\label{tab:research_problems}
\resizebox{\textwidth}{!}{%
\begin{tabular}{|c|c|l|l|l|}
\hline
\rowcolor[HTML]{C0C0C0}
\multicolumn{1}{|c|}{\textbf{Application}}&
  \multicolumn{1}{|c|}{\textbf{Ref.}}  &
  \multicolumn{1}{|c|}{\textbf{Target issue}} &
  \multicolumn{1}{|c|}{\begin{tabular}[c]{@{}c@{}}\textbf{Solution proposed}\\ \textit{and/or blockchain usage}\end{tabular}} &
  \multicolumn{1}{|c|}{\begin{tabular}[c]{@{}c@{}}\textbf{Supporting techniques}\\ \textit{and/or smart contract} \\ \textit{(SC) usage}\end{tabular}} \\ \hline
& \cite{chen2020blockchain, chen2019blockchain}  &
  \begin{tabular}[|c|]{@{}l@{}} Securing communication\end{tabular} &
  \begin{tabular}[|c|]{@{}l@{}} • Novel cryptographic primitive: blockchain-based proxy re-encryption. \\ • Combines proxy re-encryption, searchable encrytion, and blockchain. \end{tabular} &
  \begin{tabular}[|c|]{@{}l@{}} SC performs ciphertext \\ matching for data searches  \end{tabular}
   \\ \cmidrule{2-5}
& \cite{lu2020blockchain}  &
  \begin{tabular}[|c|]{@{}l@{}} Reliability and efficiency \\ of data sharing \end{tabular} &
  \begin{tabular}[|c|]{@{}l@{}} • Federated learning to fulfill data sharing requests correctly. \\ • Local DAG for storing shared update models, global permissioned \\\hspace{1mm} blockchain for managing data sharing requests.\end{tabular} &
  \begin{tabular}[|c|]{@{}l@{}} Federated learning, \\ DAGs \end{tabular}
   \\ \cmidrule{2-5}
\begin{tabular}[|c|]{@{}l@{}} Data trading/\\ sharing \end{tabular}&
  \cite{liu2019novel}  &    
  \begin{tabular}[|c|]{@{}l@{}}Transaction delays and \\ cold start problems\end{tabular} &
  \begin{tabular}[|c|]{@{}l@{}}• Vehicles buy and sell data. To allow new users to participate even \\\hspace{1mm} with empty accounts, this is formulated as a debt-credit system.\end{tabular} & \begin{tabular}[|c|]{@{}l@{}} Pricing strategy modelled as \\ a two-stage Stackelberg \\ game\end{tabular}
   \\ \cmidrule{2-5} 
& \cite{ahmad2019realization}  &
  \begin{tabular}[|c|]{@{}l@{}} NDN data sharing\end{tabular} &
  \begin{tabular}[|c|]{@{}l@{}} • Layered model, with NDN routers interfacing with blockchain.\end{tabular} &
  \begin{tabular}[|c|]{@{}l@{}} Named Data \\ Networking (NDN) \end{tabular}
   \\ \cmidrule{2-5}
& \cite{yang2020ldv}  &
  \begin{tabular}[|c|]{@{}l@{}} Data sharing in VSNs \end{tabular} &
  \begin{tabular}[|c|]{@{}l@{}} • Directed Acrylic Graph (DAG) based blockchain stores data within \\\hspace{1mm} relevant topic groups in VSNs. \end{tabular} &
  \begin{tabular}[|c|]{@{}l@{}} Directed Acrylic Graphs\end{tabular}
   \\ \hline
& \cite{chen2019smart} &
  \begin{tabular}[|c|]{@{}l@{}}Traffic management, \\ Autonomous driving\end{tabular} &
  \begin{tabular}[|c|]{@{}l@{}}• Vehicle platooning based on path matching. \\ • Platoon heads, chosen rotationally by reputation, pay attention to the \\\hspace{1mm} road while platoon members can relax. \end{tabular} &
  \begin{tabular}[|c|]{@{}l@{}}Platoon heads paid in \\ cryptocurrency by \\ members via SCs \end{tabular} \\ \cmidrule{2-5}
& \cite{baza2019b}  &
  \begin{tabular}[|c|]{@{}l@{}} Ride sharing\end{tabular} &
  \begin{tabular}[|c|]{@{}l@{}} • Rider makes a time-locked deposit and provides a set of obfuscated \\\hspace{1mm} locations; driver also makes a deposit until proof of pick-up. \end{tabular} &
  \begin{tabular}[|c|]{@{}l@{}} SCs prevent fraud\end{tabular}
   \\ \cmidrule{2-5}
& \cite{fu2020autonomous}  &
  \begin{tabular}[|c|]{@{}l@{}} Lane changing in \\ autonomous vehicles\end{tabular} &
  \begin{tabular}[|c|]{@{}l@{}} • Lane changing modelled as a Deep Reinforcement Learning problem. \\ • Secure collective learning framework using blockchain. \end{tabular} &
  \begin{tabular}[|c|]{@{}l@{}} Deep Reinforcement Learning (DRL)\end{tabular}
   \\ \cmidrule{2-5}
Transportation &
  \cite{song2020blockchain, li2020vehicle} &
  \begin{tabular}[|c|]{@{}l@{}}Correcting errors in \\ GPS positioning\end{tabular} &
  \begin{tabular}[|c|]{@{}l@{}}• LIDAR aided vehicles train a DNN and share positioning error \\\hspace{1mm} information with other vehicles through blockchain.\end{tabular} &
  \begin{tabular}[|c|]{@{}l@{}}SCs ensure accuracy \\ of shared models\end{tabular} \\ \cmidrule{2-5}
& \cite{zhang2020bsfp}  &
  \begin{tabular}[|c|]{@{}l@{}} Smart parking\end{tabular} &
  \begin{tabular}[|c|]{@{}l@{}} • Parking owners rent out space using blockchain. \end{tabular} &
  \begin{tabular}[|c|]{@{}l@{}} SCs realise fairness; \\ matching, advance payment \end{tabular}
   \\ \cmidrule{2-5}
& \cite{li2018efficient}  &
  \begin{tabular}[|c|]{@{}l@{}} Carpooling\end{tabular} &
  \begin{tabular}[|c|]{@{}l@{}} • Fog computing to match user carpooling requests with \\\hspace{1mm} potential drivers. \\ • Blockchain stores records, with conditional privacy. \end{tabular} &
  \begin{tabular}[|c|]{@{}l@{}} Bloom filters for \\ location anonymity\end{tabular}
   \\ \cmidrule{2-5}
& \cite{cheng2019sctsc} &
  \begin{tabular}[|c|]{@{}l@{}}Road congestion, \\ inefficiency\end{tabular} &
  \begin{tabular}[|c|]{@{}l@{}}• Blockchain holds travel related information.\\ • Degree of availability of information on the blockchain is based on \\\hspace{1mm} attributes of a vehicle, like direction of travel.\end{tabular} &
  \begin{tabular}[|c|]{@{}l@{}}CP-ABE encryption used \\ instead of ordinary PKCs.\end{tabular} \\ \hline
  & \cite{yao2019bla} & 
\begin{tabular}[|c|]{@{}l@{}}Cross datacenter \\ authentication in \\ fog computing scenario \end{tabular} &
  \begin{tabular}[|c|]{@{}l@{}}• Custom privacy preserving authentication scheme for fog computing; \\\hspace{1mm} easy re-authentication across different locations. \\ • Consortium blockchain stores authentication records of vehicles.\end{tabular} &
  \begin{tabular}[|c|]{@{}l@{}} \hspace{13mm} - \end{tabular} \\ \cmidrule{2-5}
& \cite{lin2020bcppa}  &
  \begin{tabular}[|c|]{@{}l@{}} Conditional privacy - \\ preserving authentication \end{tabular} &
  \begin{tabular}[|c|]{@{}l@{}} • Blockchain used to store certificates as transactions. \\ • Messages contain transaction ID that authenticated sender vehicle.\end{tabular} &
  \begin{tabular}[|c|]{@{}l@{}}SCs are used to broadcast \\ certificates to the blockchain\end{tabular}
   \\ \cmidrule{2-5}
& \cite{liu2020blockchain}  &
  \begin{tabular}[|c|]{@{}l@{}} Performance bottlenecks \end{tabular} &
  \begin{tabular}[|c|]{@{}l@{}} • Edge computing proxy vehicles that authenticate  vehicle groups.\end{tabular} &
  \begin{tabular}[|c|]{@{}l@{}}Proxy authentication\end{tabular}
   \\ \cmidrule{2-5}
Authentication &
   \cite{vangala2020blockchain} &
  \begin{tabular}[|c|]{@{}l@{}}Vehicle authentication for \\ accident detection\end{tabular} &
  \begin{tabular}[|c|]{@{}l@{}}• Custom certificate-based authentication scheme. \\ • Blockchain holds transactions for accident related information.\end{tabular} &
  Dynamic clustering \\ \cmidrule{2-5}
& \cite{shen2020blockchain}  &
  \begin{tabular}[|c|]{@{}l@{}} Lightweight CA for \\ location based services\end{tabular} &
  \begin{tabular}[|c|]{@{}l@{}} • Threshold proxy scheme is employed by CAs that play role of \\\hspace{1mm} distributed nodes inside a consortium type blockchain.\end{tabular} &
  \begin{tabular}[|c|]{@{}l@{}} Threshold proxy signature\end{tabular}
   \\ \cmidrule{2-5}
& \cite{tan2019secure}  &
  \begin{tabular}[|c|]{@{}l@{}}Batch Authentication + \\ Key Management\end{tabular} &
  \begin{tabular}[|c|]{@{}l@{}}• Certificateless auth; TA and OBU establish session keys with \\\hspace{1mm} operations performed explicitly at the cloud side. \\ • Group key generation for efficient and secure V2V communication.\end{tabular} &
  \begin{tabular}[|c|]{@{}l@{}}Group key\end{tabular}
   \\ \hline
&
  \begin{tabular}[|c|]{@{}l@{}}\cite{li2019iterative, su2018secure} \\ \cite{wang2019bsis, li2020consortium} \end{tabular}  &
  \begin{tabular}[|c|]{@{}l@{}}Coordinating \\ charging-discharging \\ schedules of vehicles\end{tabular} &
  \begin{tabular}[|c|]{@{}l@{}}• Aggregators formulate a schedule  based on requests sent as \\\hspace{1mm} blockchain transactions by EVs.\\ • The schedule is formulated as an optimization problem.\end{tabular} & \begin{tabular}[|c|]{@{}l@{}} SCs set prices, \\ maximise utility \end{tabular} \\ \cline{2-5}
Smart grid  & \cite{li2020privacy} &
  \begin{tabular}[|c|]{@{}l@{}} Charging services \\ with focus on privacy\end{tabular} &
  \begin{tabular}[|c|]{@{}l@{}}• Blockchain with fog computing to reduce latency.\\ • Selective storage of only sensitive data in the blockchain. \end{tabular} &
  Fog Computing \\ \cline{2-5}
 & \cite{wang2019bbars} &
  \begin{tabular}[|c|]{@{}l@{}}Anonymously rewarding \\ vehicles for selling energy\end{tabular} &
  \begin{tabular}[|c|]{@{}l@{}}• Exchange of energy happens for appropriate (secure) payment of \\\hspace{1mm} blockchain cryptocurrency.\end{tabular} &
  \begin{tabular}[|c|]{@{}l@{}}SCs decide remuneration\end{tabular} \\ \cmidrule{2-5}
 & \cite{zhou2019blockchain} &
  \begin{tabular}[|c|]{@{}l@{}} Complete energy \\ trading framework\end{tabular} &
  \begin{tabular}[|c|]{@{}l@{}}• Discharging EVs compensated by charging EVs with cryptocurrency. \\ • Incentive compatible Demand-Response paradigm is used.\end{tabular} &
  \begin{tabular}[|c|]{@{}l@{}} SCs maximise social \\ welfare, in terms of \\ revenue generated.\end{tabular} \\ \hline

& \cite{singh2020blocked}  &
  \begin{tabular}[c]{@{}l@{}} Edge-based data processing \\ framework in VANETs\end{tabular} &
  \begin{tabular}[c]{@{}l@{}}• Tasks allocated to containers on edge nodes, based on time and \\\hspace{1mm} resources needed. Formulated as multi-objective optimization problem. \\ • Containers can be migrated to other edge nodes using blockchain.\end{tabular} &
  \begin{tabular}[c]{@{}l@{}} Containerization \end{tabular}
   \\ \cmidrule{2-5}
\begin{tabular}[c]{@{}l@{}} Resource sharing\end{tabular} & 
  \begin{tabular}[c]{@{}l@{}} \cite{chai2019proof, lin2020blockchain} \\ \cite{wang2020consortium}  \end{tabular}&
  \begin{tabular}[c]{@{}l@{}} Complete architecture \\ for resource sharing\end{tabular} &
  \begin{tabular}[c]{@{}l@{}} • Spectrum and computation resources paid for with cryptocurrency. \end{tabular} &
  \begin{tabular}[c]{@{}l@{}} SCs determine pricing by \\ matching demand and supply \end{tabular}
   \\ \cmidrule{2-5}
& \cite{huang2020securing}  &
  \begin{tabular}[c]{@{}l@{}} Vehicular fog computing\\with parked vehicles\end{tabular} &
  \begin{tabular}[c]{@{}l@{}} • Requester uses blockchain currency to pay vehicles for using \\\hspace{1mm} their computation resources. \\• Problem formulated as two-stage Stackelberg game.\end{tabular} &
  \begin{tabular}[c]{@{}l@{}} SCs mediate requesters \\ and performers/providers\end{tabular}
   \\ \cmidrule{2-5}
& \cite{yao2019lightweight}  &
  \begin{tabular}[c]{@{}l@{}} IDaaS with Vehicular \\ cloud computing \end{tabular} &
  \begin{tabular}[c]{@{}l@{}}• Identity-as-a-Service model for vehicles and vehicular clouds.\\ • Personally identifiable information encrypted and stored in blockchain. \end{tabular} &
  \begin{tabular}[c]{@{}l@{}} Encryption with CP-ABE\end{tabular}
   \\ \hline
& \cite{wang2020blockchain}  &
  \begin{tabular}[c]{@{}l@{}} Vehicle cooperation \\ for crowdsensing tasks\end{tabular} &
  \begin{tabular}[c]{@{}l@{}}• Vehicle team selection and payment method based on blockchain. \\ • Credit score determined by number of successful completions. \end{tabular} &
  \begin{tabular}[c]{@{}l@{}} Reverse auction \end{tabular}
   \\ \cmidrule{2-5}
\begin{tabular}[c]{@{}l@{}} Crowdsourcing / \\ Crowdsensing \end{tabular}&
  \cite{lai2019spir}  &
  \begin{tabular}[c]{@{}l@{}}Real time map updates\end{tabular} &
  \begin{tabular}[c]{@{}l@{}}• Blockchain based credit management system --- a privacy \\\hspace{1mm} preserving incentive mechanism.\end{tabular} & \begin{tabular}[c]{@{}l@{}}  optimization problem, \\ reverse auction mechanism \end{tabular}
   \\ \cmidrule{2-5}
& \cite{zhang2020decentralized}  &
  \begin{tabular}[c]{@{}l@{}} Location privacy in \\ crowdsourcing tasks\end{tabular} &
  \begin{tabular}[c]{@{}l@{}} • Area grid recursively partitioned using quad tree function. \\ • Workers share location data over blockchain; recursive partitioning \\\hspace{1mm} allows selection of privacy levels. Task requesters access blockchain.\end{tabular} &
  \begin{tabular}[c]{@{}l@{}} \hspace{13mm} --- \end{tabular}
   \\ \hline
& \cite{zhang2019blockchain}  &
  \begin{tabular}[c]{@{}l@{}} Vehicular SDN \end{tabular} &
  \begin{tabular}[c]{@{}l@{}} • Blockchain used to manage the network commands \\\hspace{1mm} for control plane securely.\end{tabular} &
  \begin{tabular}[c]{@{}l@{}} Q-learning to manage \\ system state \end{tabular}
   \\ \cmidrule{2-5}
&
  \cite{sharma2018energy} &
  \begin{tabular}[c]{@{}l@{}}Large energy consumption \\ in Blockchain enabled IoV\end{tabular} &
  \begin{tabular}[c]{@{}l@{}}• Model that manages energy consumed for consensus by selectively \\\hspace{1mm} representing some nodes by their associated cluster head.\end{tabular} &
  Distributed Clustering \\ \cmidrule{2-5}
\begin{tabular}[c]{@{}l@{}}BIoV\\architecture\end{tabular} & \cite{gao2019blockchain} & \begin{tabular}[c]{@{}l@{}} Performance and security /\\ trust services in VANETs \end{tabular} & \begin{tabular}[c]{@{}l@{}} • Architecture for VANETs that combine SDN, blockchain, and \\\hspace{1mm} fog computing technologies. \\ • Blockchain provides secure communication. \end{tabular} & \hspace{13mm} --- \\ \cmidrule{2-5}
& \cite{jameel2020efficient}  &
  \begin{tabular}[c]{@{}l@{}} Mining cluster \\ selection\end{tabular} &
  \begin{tabular}[c]{@{}l@{}} • Offloading vehicles and mining clusters are matched based on \\\hspace{1mm} (1) transmission rate and (2) available cluster resources for mining\end{tabular} &
  \begin{tabular}[c]{@{}l@{}} \hspace{13mm} ---\end{tabular}
   \\ \cmidrule{2-5}
& \cite{lei2017blockchain}  &
  \begin{tabular}[c]{@{}l@{}} Key management\end{tabular} &
  \begin{tabular}[c]{@{}l@{}} • Traditional architecture: Different CAs maintain identity information \\\hspace{1mm} for different regions; crossovers involve inter-CA communication. \\• Proposed architecture: CAs replaced with a blockchain network. \end{tabular} &
  \begin{tabular}[c]{@{}l@{}} \hspace{13mm} --- \end{tabular}
   \\ \hline

\end{tabular}
}
\end{table*}

\subsection{Transportation}
This application area deals primarily with vehicle movement and management. Coordinating vehicles in real-time allows for more effective movement - one might think of how traffic is managed and coordinated even now with traffic lights and signs, and consider how that concept could be extended digitally with extremely specific travel information of vehicles \cite{hassija2020traffic}. It opens up possibilities for collaboration that are either too centralized or too computation-heavy for anything other than a distributed system. Some of these possibilities are outlined below:

\begin{itemize}
    \item \textbf{Ride sharing/Carpooling:} Reducing the number of vehicles on the road is a necessary step towards better environmental conditions and road safety. Two aspects of blockchain, namely, built-in cryptocurrency and smart contracts make it suitable for this task.
    \item \textbf{Platooning:} The concept of cars forming a group and navigating as a group has a couple of advantages. Firstly, it reduces traffic congestion since coordinating a few groups is easier than coordinating several vehicles, and secondly, it reduces the chances of accidents. Blockchain is used in \cite{chen2019smart} as a transaction framework that achieves this.
    \item \textbf{Smart parking:} Many parking lots are used inefficiently, with a fairly simple parking model involving a flat and hourly rate. These rates are heavily dependent on local demand. That demand could be spread over a larger area, which reduces the number of underutilized parking areas. Conceptually, it is just more efficient resource allocation. Many studies have \textcolor{black}{focused} on developing intelligent parking mechanisms \cite{chamola2020iot,hassija2020parking}, for example Zhang et al. proposed a blockchain-based smart parking framework that links customers to available parking lots \textcolor{black}{\cite{zhang2020bsfp}.}
\end{itemize}

\begin{figure*}
    \centering
    \includegraphics[width =2 \columnwidth]{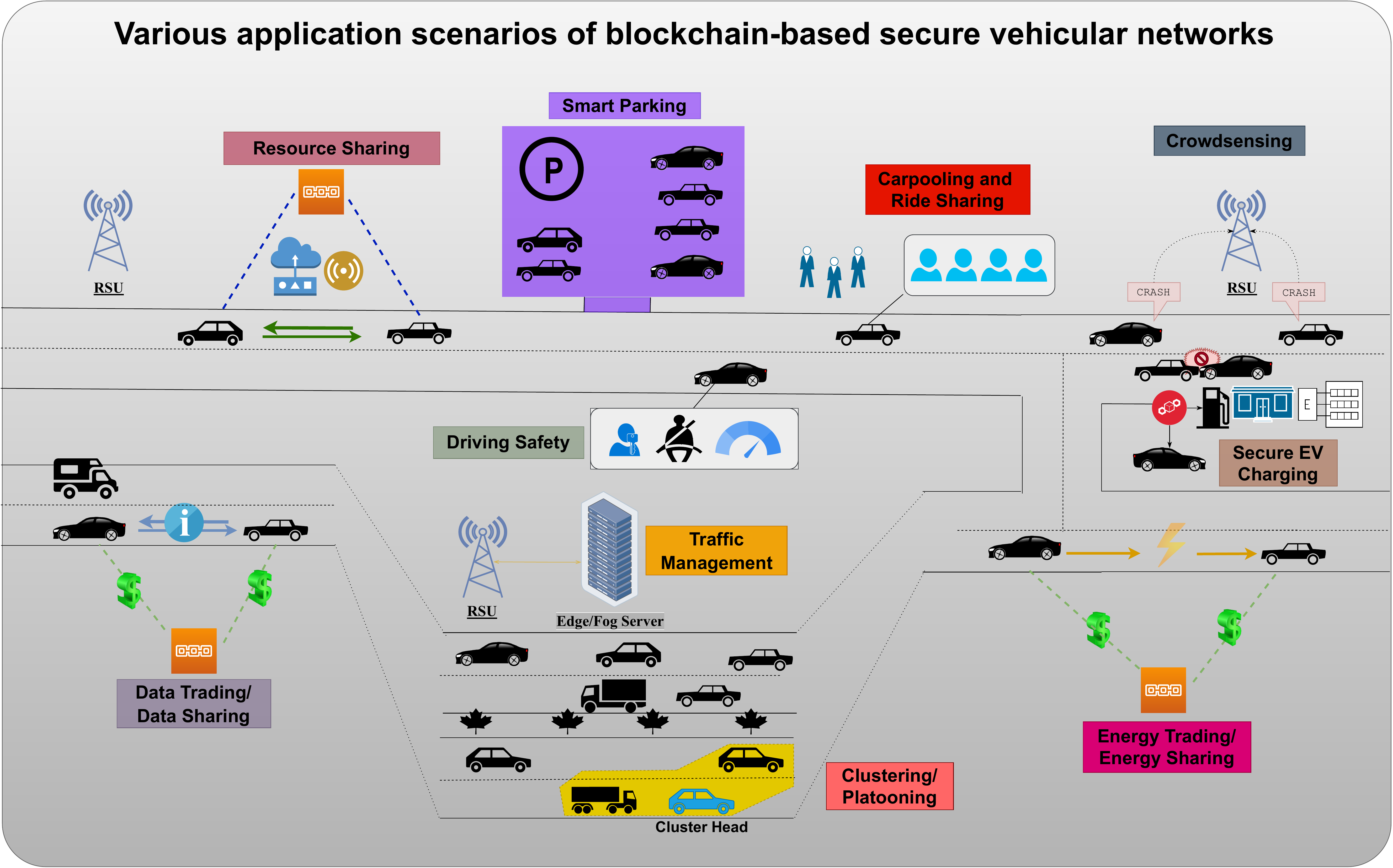}
    \caption{Thematic view of the various application scenarios in which blockchain is used for ensuring security.}
    \label{fig:appnetsc}
\end{figure*}

\subsection{Smart Grid}
This application area deals with EVs \cite{miglani2020blockchain}. The key feature of smart grids that sets them apart from traditional grids is the applicability of big data analytics. Consumers may also sell energy to the grid, which means extra electricity need not be generated needlessly if demand and supply can be met with the existing resources. Smart grids need to be fault-tolerant with the ability to quickly handle any possible faults. Again, distributed systems come to their rescue. Scalability is a major challenge in this sort of application; since the infrastructure and power distribution complexity scale very disproportionately to the geographic area. Newer blockchain architectures are being proposed to provide scalability \cite{hassija2020blockchain}. Secure EV charging is the most common application for smart grids. With the advent of electric vehicles, coordinating charging schedules to balance out supply and demand over time in a way that maintains security is the primary focus of these kinds of research works. Blockchain also provides a payment platform for rewarding vehicles that share their energy to the grid, for example, if someone realizes they have excess charge left over after making a trip.

\subsection{Resource \textcolor{black}{Sharing}}
Cloud computing is the key driving technology of this research area. The paradigm is the same - exchange of computational resources for payment, except with the complexities of vehicular networks and edge computing. Blockchain can be used to construct a distributed open market type of system, rather than rely on computation provided by a single third party. It is assumed that vehicles, unlike typical embedded systems, may be equipped with relatively higher processing capabilities, which allows them to be computational resource providers; they may also behave as consumers, purchasing resources from the RSUs. 

\subsection{Crowdsensing/Crowdsharing}
Similar to data sharing, crowdsensing applications focus more on optimizing the system for aggregated data collection rather than point-to-point data trading. Crowdsensing allows service providers to gather data in real-time, and blockchain allows people to contribute (maybe for payment) securely. This is useful especially in the map and location-based services, for example, notifying vehicles of a crash up ahead or road closure. Some form of crowdsourcing is already being practised currently, but it lacks a strong incentive system which limits its usefulness.

\subsection{BIoV Architecture}
Every application area analyzed so far has been some integration of blockchain and IoV technologies. But this research area deserves to be mentioned separately since it is slightly different in terms of research focus. Much research work has gone into optimizing the core features of blockchain to make it more adaptable to VANETs. For example, \cite{hu2019blockchain} proposed a custom consensus algorithm for vehicular networks, and \cite{sharma2018energy} introduced an energy-efficient clustering protocol specifically for blockchain-enabled vehicular networks. This research direction is conceptually distinct from the others in that the primary problem being addressed is not that of providing a service through blockchain use, but that of overcoming existing limitations in blockchain and/or VANETs by changing some of their core features.

\subsubsection*{\textcolor{black}{\textbf{Summary}}}
\textcolor{black}{In this section, several blockchain-based IoV frameworks were categorised for different application areas. Scenarios like data trading and sharing, transportation, smart grid, resource sharing, crowdsensing/crowdsourcing and BIoV architecture are discussed in detail, explaining their utility, and each of the frameworks is categorised into one of these groups.}

\begin{table*}[]
\centering
\caption{Summary of various security requirements, a few frameworks meeting those requirements, their definitions, and attacks they help to mitigate.}
\label{tab:security_req_vs_attacks}
\resizebox{\textwidth}{!}{%
\begin{tabular}{|c|c|c|c|c|l|c|}
\hline
{\color[HTML]{C0C0C0} } &
  \multicolumn{2}{c|}{\cellcolor[HTML]{C0C0C0}\textbf{\begin{tabular}[c]{@{}c@{}}Security \\ requirements\end{tabular}}} &
  \cellcolor[HTML]{C0C0C0}\textbf{\begin{tabular}[c]{@{}c@{}}Alternate names / \\ Other similar \\ security requirements\end{tabular}} & \cellcolor[HTML]{C0C0C0}\textcolor{black}{\textbf{References}} &
  \multicolumn{1}{c|}{\cellcolor[HTML]{C0C0C0}\textbf{Definition}} &
  \cellcolor[HTML]{C0C0C0}\textbf{Attacks covered} \\ \hline
\cellcolor[HTML]{C0C0C0} &
  \multicolumn{2}{c|}{ \textit{Decentralization}} &
  \begin{tabular}[c]{@{}c@{}}Without third\\  party reliance\end{tabular} & \begin{tabular}[c]{@{}c@{}} \textcolor{black}{\cite{liu2019novel, shi2019dynamic, kang2017enabling}}, \\ \textcolor{black}{\cite{sheikh2019secured}}\end{tabular}&
  \begin{tabular}[c]{@{}l@{}}Nodes participate in a P2P\\ network eliminating the need\\ of a third party.\end{tabular} &
  SPoF (Single Point of Failure) \\ \cline{2-7} 
\cellcolor[HTML]{C0C0C0} &
  \multicolumn{2}{c|}{ \textit{\begin{tabular}[c]{@{}c@{}}Traceability\\ (Via transparency )\end{tabular}}} &
  Transparency & \textcolor{black}{\cite{zhang2019data,li2019toward}} &
  \begin{tabular}[c]{@{}l@{}}All the events in the network\\ can be sequentially traced.\end{tabular} &
  Repudiation attack \\ \cline{2-7} 
\cellcolor[HTML]{C0C0C0} &
  \multicolumn{2}{c|}{ \textit{Tamper proof}} &
  Data integrity & \textcolor{black}{\cite{li2019toward, zhang2019data}} &
  \begin{tabular}[c]{@{}l@{}}The stored data inside the\\ network cannot be tampered\\ with.\end{tabular} &
  \begin{tabular}[c]{@{}c@{}}Data tampering/modification attack\\  MITM \\ Block tampering attack\end{tabular} \\ \cline{2-7} 
\cellcolor[HTML]{C0C0C0} &
  \multicolumn{2}{c|}{ \textit{Unforgeability}} &
  - & \textcolor{black}{\cite{kang2017enabling, li2019toward}} &
  \begin{tabular}[c]{@{}l@{}}Adversaries cannot forge data\\ or user's digital signature.\end{tabular} &
  Forgery Attack \\ \cline{2-7} 
\multirow{-10}{*}{\cellcolor[HTML]{C0C0C0}\textbf{\begin{tabular}[c]{@{}c@{}}BLOCKCHAIN \\ SPECIFIC\end{tabular}}} &
  \multicolumn{2}{c|}{ \textit{Consensus mehanism}} &
  Public audit & \textcolor{black}{\cite{zhou2019secure,liu2019novel}} &
  \begin{tabular}[c]{@{}l@{}}Data stored in the network is\\ verified by the network nodes.\end{tabular} &
  MITM \\ \hline
\cellcolor[HTML]{C0C0C0} &
  \multicolumn{2}{c|}{ \textit{Non-repudiation}} &
  - & \begin{tabular}[c]{@{}l@{}} \textcolor{black}{\cite{zhang2019data, li2019iterative, li2019toward}}, \\ \textcolor{black}{\cite{cheng2019sctsc, yao2019bla}} \end{tabular}&
  \begin{tabular}[c]{@{}l@{}}Ensures that senders cannot\\ deny that they have sent\\ messages or made transactions.\end{tabular} &
  \begin{tabular}[c]{@{}c@{}}Repudiation attack,\\  Masquerade/impersonation attack,\\  Forgery attack\end{tabular} \\ \cline{2-7} 
\cellcolor[HTML]{C0C0C0} &
    &
   \textit{Anonymity} &
  Untraceability & \begin{tabular}[c]{@{}l@{}} \textcolor{black}{\cite{zheng2019traceable, zhang2019data, huang2018lnsc}}, \\ \textcolor{black}{\cite{li2019iterative, wang2019bbars}} \end{tabular} &
  \begin{tabular}[c]{@{}l@{}}Ensures that personal details of\\ users are not compromised while\\ still allowing them to participate\\ fully.\end{tabular} &
  Tracking attack \\ \cline{3-7} 
\cellcolor[HTML]{C0C0C0} &
    &
    &
  \begin{tabular}[c]{@{}c@{}}Forward and \\ backward security\end{tabular} & \begin{tabular}[c]{@{}l@{}} \\ \\ \\ \textcolor{black}{\cite{huang2018lnsc,wang2019bbars,zhou2019secure}}, \\ \cite{cheng2019sctsc,li2020privacy} \end{tabular} &
  \begin{tabular}[c]{@{}l@{}}Ensuring that attackers cannot\\ use the relation between two\\ messages sent by the same user\\ to violate that user’s privacy.\end{tabular} &
  Background analysis attack \\ \cline{6-7} 
\cellcolor[HTML]{C0C0C0} &
  \multirow{-3}{*}{ \textit{\begin{tabular}[c]{@{}c@{}}Privacy \\ preservation\end{tabular}}} &
  \multirow{-2}{*}{ \textit{Unlinkability}} &
  Known key security & &
  \begin{tabular}[c]{@{}l@{}}Privacy of future sessions is \\ guaranteed even in case the\\ session key is leaked for one\\ of the previous sessions.\end{tabular} &
  Ephemeral Secret Leakage (ESL) \\ \cline{2-7} 
\cellcolor[HTML]{C0C0C0} &
  \multicolumn{2}{c|}{ \textit{\begin{tabular}[c]{@{}c@{}}Traceability\\  (via conditional privacy)\end{tabular}}} &
  \begin{tabular}[c]{@{}c@{}}Conditional \\ linkability\end{tabular} & \begin{tabular}[c]{@{}l@{}} \textcolor{black}{\cite{zheng2019traceable,li2019toward,cheng2019sctsc}}, \\ \textcolor{black}{\cite{gao2019blockchain,yao2019bla}} \end{tabular} &
  \begin{tabular}[c]{@{}l@{}}In the event of an attack, the\\ identity of the offending vehicle\\ must be revealed to authorities.\end{tabular} &
  \begin{tabular}[c]{@{}c@{}}Double claim attack\\ Repudiation attack\end{tabular} \\ \cline{2-7} 
\cellcolor[HTML]{C0C0C0} &
  \multicolumn{2}{c|}{ \textit{Wallet security}} &
  Account security & \begin{tabular}[c]{@{}l@{}} \textcolor{black}{\cite{li2019toward,kang2017enabling,liu2019electric}}, \\ \textcolor{black}{\cite{chen2019toward, lin2020blockchain}} \end{tabular} &
  \begin{tabular}[c]{@{}l@{}}The security of the e-wallet\\ used by the vehicles is\\ guaranteed by the network.\end{tabular} &
  - \\ \cline{2-7} 
\cellcolor[HTML]{C0C0C0} &
  \multicolumn{2}{c|}{ \textit{Scalability}} &
  - & \begin{tabular}[c]{@{}l@{}} \textcolor{black}{\cite{zhang2019data,li2019iterative,sheikh2019secured}}, \\ \textcolor{black}{\cite{javaid2020scalable,feng2019bpas}} \end{tabular} &
  \begin{tabular}[c]{@{}l@{}}The network functionality and \\ performance are not affected\\ when the number of\\ participating vehicles become\\ very large.\end{tabular} &
  - \\ \cline{2-7}
\multirow{-24}{*}{\cellcolor[HTML]{C0C0C0}\textbf{\begin{tabular}[c]{@{}c@{}}NETWORK \\ SPECIFIC\end{tabular}}} &
  \multicolumn{2}{c|}{ \textit{\begin{tabular}[c]{@{}c@{}}Low latency and \\ high throughput\end{tabular}}} &
  \begin{tabular}[c]{@{}c@{}}Low computational \\ overhead\end{tabular} & \textcolor{black}{\cite{ferrag2019deepcoin}} &
  \begin{tabular}[c]{@{}l@{}}There are minimal network\\ delays and the network is\\ capable of adding a high\\ number of transactions to\\ blockchain in a short time span.\end{tabular} &
  \begin{tabular}[c]{@{}c@{}}All kinds of attacks \\ ( as more latency gives more window\\  to adversaries for carrying out \\ attacks)\end{tabular} \\ \hline
\end{tabular}
}
\end{table*}

\section{Categorisation based on Security Features}
\label{sec:security_perspective}
\color{blue}
The security perspective section categorises different blockchain-based security works based on the following category types.
\begin{enumerate}[i.]
    \item Security requirements met
    \item Protection against network-specific security attacks
    \item Authentication techniques
    \item Security proof
\end{enumerate}
\color{black}
The purpose and scope of each category and the definition of fields within them have been defined in the subsequent sections. This will give the readers a holistic view of the research done in vehicular network security using blockchain-based frameworks, and thus help in further research in this area.

\subsection{Security Requirements}
\label{sec:security_requirements}
A security requirement is a condition which a network should meet to mitigate attacks and vulnerabilities in the network. There have been various studies that have defined different security requirements for vehicular networks \cite{manvi2017survey,raya2005security,raya2007securing}. Blockchain technology implicitly provides some of these security requirements due to its decentralized nature and tamper-proof storage mechanism. Thus, any blockchain-based security framework will meet the following security requirements by default.
\begin{enumerate}
    \item \textbf{Decentralization:} Blockchain eliminates the involvement of any third party by enabling a P2P network where some of the blockchain nodes verify the transactions. This preserves the privacy of the vehicles by eliminating the need for sharing their details with a third party \cite{liu2019novel}. This property has several applications in vehicular networks such as decentralized communication \cite{shi2019dynamic}, data sharing \cite{kang2017enabling}, and identity management \cite{sheikh2019secured} among others.
    
    \item \textbf{Tamper resistance:} The data recorded in the blockchain is difficult to tamper with because it is organised in the form of special structures such as hash chain \cite{li2019toward}, and every block contains the hash of the previous block. This ensures irreversibility and immutability as tampering with the data in any block will change its hash value and will disconnect it from the blockchain. Also, blockchain's distributed nature ensures that the data has not been tampered with at any intermediary stages because of elimination of the third party, i.e., decentralization also leads to tamper resistance \cite{zhang2019data}.

    \item \textbf{Unforgeability:} It refers to the ability of a network to resist adversaries from forging data or a user's digital signature. The decentralized nature of the blockchain combined with its digitally signed transactions guarantee this and ensure that adversaries are not able to pose as other users \cite{kang2017enabling}. In \cite{li2019toward}, Li et al. use a blockchain-based fair and anonymous ad dissemination scheme to ensure unforgeability using a popular authentication algorithm called Zero-Knowledge Proof.
   
    \item \textbf{Traceability (via cryptographic hash):} Each block in the blockchain contains the cryptographic hash of the previous block, thus ensuring traceability \cite{zhang2019data,li2019toward}. Each node can trace and verify the correspondence of the data. This will help in tracing any malicious activity or message circulation thus avoiding confusion and accidents in the vehicular network.
    
% \begin{figure*}
%         \centerline{\includegraphics[width =1.5 \columnwidth]{Figures/5_Secuirty_perspective_final.pdf}}
%         \caption{Overview of the security perspective section.}
%         \label{fig:secuirty_perspective}
% \end{figure*}

    \item \textbf{Public audit:} Blockchain helps in implementing public audits via its consensus mechanism. The block created by the miners must satisfy the criteria of the consensus mechanism used and should also be independently verified by other nodes in the network. This feature has been used in \cite{zhou2019secure} for authentication purposes in the vehicle to grid energy trading, and in \cite{liu2019novel} for publicly auditing transnational data trading in the IoV environment.
\end{enumerate}

Apart from the above security requirements, there are other security requirements that a blockchain-based scheme must satisfy to increase its robustness. In this study, the following is the list of security requirements considered (excluding the ones implicitly provided by blockchain):

\begin{enumerate}
    \item \textbf{Non-repudiation:} It ensures senders cannot deny the transmission of a message and also ensures easy identification of the vehicle nodes in case of accidents. Mitigation of repudiation \textcolor{black}{attacks} comes under this security requirement. It can be met by ensuring the following.
    
    \begin{enumerate}
        \item All transmitted messages are signed by a transmitting node via its anonymous public key to ensure that the node cannot claim to be some other node in the network (which can result in a masquerade attack).  
        \item The node cannot claim that the message was replayed since the message is timestamped. It is crucial for investigation agencies for finding the chain of events and message details in case of a mishap.
        \item Vehicles cannot falsify their location information due to the implementation of secure positioning solutions.
    \end{enumerate}

    \item \textbf{Privacy preservation:} Frameworks meeting this security requirement ensures that the private information of the participating nodes is not disclosed to the public or malicious parties. Also, the privacy of the drivers should be guaranteed against unauthorized observers \cite{li2018creditcoin}. It has mainly two important features:
      \begin{enumerate}
          \item \textbf{Anonymity:} Specific personal details of the nodes such as name and vehicle type should not be disclosed. It prevents malicious parties from tracking the activities of a user by ensuring that for each incoming transaction, all possible senders are equiprobable, i.e., untraceablility \cite{wang2019bbars} (not to be confused with traceability as a security requirement). Anonymity is generally achieved by using pseudonyms and cryptographic techniques. It leads to mitigation of tracking attacks \cite{zheng2019traceable}.
          \item \textbf{Unlinkability:} It ensures that the attackers cannot link received \textcolor{black}{messages} sent from the same sender; or in other words, for any two outgoing transactions, it is impossible to prove that they were sent to the same person \cite{wang2019bbars}. This is similar to forward security \cite{liu2020blockchain}.
      \end{enumerate}

\item \textbf{Traceability (via conditional privacy):} By linking a vehicle's pseudonym to its true identity, a government agency or trusted third party must be able to trace the identity of the vehicle if it is found to be malicious. \cite{li2019toward} also refer to it as conditional linkability and meet this security requirement to prevent a \say{double claim} attack.

\item \textbf{Wallet security:} Most vehicular frameworks have an incentive mechanism and reward the vehicles with cryptocurrency for behaving properly in the network. Vehicles also pay cryptocurrency to each other in exchange for services and data. Hence, very often each vehicle user has an e-wallet associated with it. Meeting this security requirement ensures that the wallet of the users is secured against malicious attacks. In \cite{li2019toward}, each user employs a client providing secure wallet services to manage the cryptocurrency tokens and connect to the blockchain network. \cite{liu2019novel} guarantees the account security of vehicles using an encrypted digital signature. Adversaries cannot open the vehicle users' wallets without the corresponding keys and certificates.

\item \textbf{Scalability:} Scalability ensures that the framework is efficient even when there are a large number of participating vehicles in the network. For a blockchain-based framework, the measure of scalability is the blockchain's ability to manage large volumes of transactions \cite{zhang2018towards}. Low computational overhead and a dynamic consensus algorithm that adapts to the traffic volume is a way to achieve this \cite{javaid2020scalable}. According to \cite{liu2019electric} decentralization also leads to scalability. To evaluate whether a framework is scalable or not, various simulation platforms can be used and the results can be compared against benchmark blockchain frameworks such as Hyperledger Caliper \cite{li2020consortium}.

\item \textbf{Low latency and high throughput:} Low latency refers to fewer network delays, and high throughput refers to the ability of the blockchain to add a large number of transactions to the chain in a limited time. These are important requirements because a slower network gives more time to the adversaries to execute attacks \cite{sheikh2019secured}. This security requirement is subjective and not very well defined but is still worth discussing considering the security of a network.
\end{enumerate}

Table \ref{tab:security_req_vs_attacks} lists out various security requirements, frameworks meeting those requirements, their corresponding definitions, and the attacks they help to mitigate.

\subsection{Network Specific Security Attacks}
\label{sec:netwrok_security_attacks}
Vehicular networks are susceptible to many kinds of security attacks because of operating in an open and dynamic environment. Hence most security schemes ensure that their proposed frameworks provide resistance to well-known network security attacks. In this section, we discuss the common network security attacks which have been tackled in the existing blockchain-based security schemes for vehicular networks. The list of security attacks is as follows.

\begin{table*}[]
\caption{\textcolor{blue}{A summary of the network specific security attacks and the blockchain-based frameworks addressing these attacks.}}
\begin{tabular}{|
>{\columncolor[HTML]{EFEFEF}}c |c|c|c|}
\hline
\cellcolor[HTML]{C0C0C0}{\color[HTML]{333333} Attack Type}                & \cellcolor[HTML]{C0C0C0}Definition                                                                                                                     & \cellcolor[HTML]{C0C0C0}Security Concerns                                                                                              & \cellcolor[HTML]{C0C0C0}\begin{tabular}[c]{@{}c@{}}Frameworks \\ Addressing the Issue\end{tabular} \\ \hline
Replay                                                                    & \begin{tabular}[c]{@{}c@{}}Imitate a legitimate user\\ and replay already \\ transmitted data\end{tabular}                                             & Authenticity, Confidentiality   &   \begin{tabular}[c]{@{}c@{}} \cite{vangala2020blockchain, zheng2019traceable, li2019iterative, cheng2019sctsc, lin2020bcppa, liu2020blockchain, tan2019secure}, \\ \cite{liu2019blockchain, sutrala2020design, kamal2020blockchain, feng2019bpas, javaid2020scalable, huang2018lnsc}       \end{tabular}   \\ 
\hline
\begin{tabular}[c]{@{}c@{}}Man-in-the-\\ middle\end{tabular}              & \begin{tabular}[c]{@{}c@{}}Adversary intercepts the\\ communication\\ between two nodes of \\ the network secretly\end{tabular}                        & \begin{tabular}[c]{@{}c@{}}Confidentiality, Physical \\ security of the IoV network  \\ resources\end{tabular}                         &   \begin{tabular}[c]{@{}c@{}}     \cite{vangala2020blockchain, li2019iterative, cheng2019sctsc, sheikh2019secured, liu2020blockchain, lin2020bcppa, chen2019secure, shen2020blockchain}, \\ \cite{sutrala2020design, huang2018lnsc, liu2019electric, ahmad2020marine}      \end{tabular}                                  \\ \hline
Impersonation                                                             & \begin{tabular}[c]{@{}c@{}}Impersonate a \\ legitimate node in the\\ network secretly\end{tabular}                                                     & Authenticity, Confidentiality                    & \begin{tabular}[c]{@{}c@{}} \cite{vangala2020blockchain, li2019iterative, li2019toward, li2020privacy, lin2020bcppa, wang2019bsis}, \\ \cite{sutrala2020design, feng2019bpas, huang2018lnsc, javaid2020scalable} \end{tabular}                                                                                                   \\ \hline
\begin{tabular}[c]{@{}c@{}}Privileged \\ Insider\end{tabular}             & \begin{tabular}[c]{@{}c@{}}Malicious insider has\\ privilege account \\ credentials/ privilege\\ access in the network\end{tabular}                    & \begin{tabular}[c]{@{}c@{}}Confidentiality, Data loss, \\ Physical security of the \\ IoV network resources\end{tabular}               &  \begin{tabular}[c]{@{}c@{}} \cite{vangala2020blockchain, ahmad2020marine}     \end{tabular}                                                                                             \\ \hline
Spoofing                                                                  & \begin{tabular}[c]{@{}c@{}}Impersonate a user/ \\ node of the network and \\ get unauthorized access\end{tabular}                                      & \begin{tabular}[c]{@{}c@{}}Authenticity, Physical security \\ of the IoV network resources\end{tabular}                                &                   \cite{rathee2020crt, sheikh2019secured, li2020vehicle, javaid2020scalable, gao2019blockchain, yang2018blockchain}        \\ \hline
\begin{tabular}[c]{@{}c@{}}Malicious \\ Third Party\end{tabular}          & \begin{tabular}[c]{@{}c@{}}Adversary generates\\ false ratings to disturb \\ the trust records\end{tabular}                                            & \begin{tabular}[c]{@{}c@{}}Authenticity, Network \\ stability\end{tabular}                                                             &  \begin{tabular}[c]{@{}c@{}}\cite{vangala2020blockchain, su2018secure, sutrala2020design, chai2019proof, wang2019bsis, yang2018blockchain}      \end{tabular}        \\ \hline
\begin{tabular}[c]{@{}c@{}}Ephemeral\\ Secret Leakage\end{tabular}        & \begin{tabular}[c]{@{}c@{}}Privacy is broken by \\ obtaining user's \\ credentials illegitimately\end{tabular}                                         & \begin{tabular}[c]{@{}c@{}}Authenticity, Confidentiality, \\ Data loss, Physical security \\ of the IoV network resources\end{tabular} &   \begin{tabular}[c]{@{}c@{}} \cite{vangala2020blockchain, huang2018lnsc, javaid2020scalable, yao2019bla, shen2020blockchain, zhou2019blockchain}, \\ \cite{liu2019blockchain} \end{tabular} \\ 
\hline
Sybil & \begin{tabular}[c]{@{}c@{}}Adversary tries to\\ control a significant\\ part of network under\\ false identities\end{tabular}
& \begin{tabular}[c]{@{}c@{}}Authenticity, Network Stability, \\Physical security of the IoV\\ network resources\end{tabular}
& \cite{cheng2019sctsc, li2020privacy, wang2019bsis, yang2020ldv} \\
\hline
\begin{tabular}[c]{@{}c@{}}Denial of\\ Service\end{tabular}               & \begin{tabular}[c]{@{}c@{}}IoV network resources\\ are made unavailable \\ by flooding the node\\ with false requests\end{tabular}                     & \begin{tabular}[c]{@{}c@{}}Authenticity, Network Stability, \\ Resource depletion\end{tabular}                                         &    \begin{tabular}[c]{@{}c@{}}    \cite{rathee2020crt, shi2019dynamic, lai2019spir}   \end{tabular}                                                                                          \\ \hline
\begin{tabular}[c]{@{}c@{}}Distributed\\ Denial of\\ Service\end{tabular} & \begin{tabular}[c]{@{}c@{}}IoV network resources \\ are made unavailable\\ across several nodes\\  by flooding them with\\ false requests\end{tabular} & \begin{tabular}[c]{@{}c@{}}Authenticity, Network Stability, \\ Resource depletion\end{tabular}                                         &    \begin{tabular}[c]{@{}c@{}}    \cite{li2019iterative, li2019toward, feng2019bpas, lin2020bcppa}                                       \end{tabular}                                                      \\ \hline
Repudiation                                                               & \begin{tabular}[c]{@{}c@{}}An entity of the network\\ falsely denies \\ participation in an \\ activity\end{tabular}                                   & Authenticity            &   \begin{tabular}[c]{@{}c@{}}           \cite{lei2017blockchain, zhang2019data, li2019iterative, cheng2019sctsc, yao2019bla, chen2019secure},  \\ \cite{li2019toward}   \end{tabular}                                                                                  \\ \hline
Eavesdropping                                                             & \begin{tabular}[c]{@{}c@{}}Secretly listening in on\\ the communication\\ between nodes\end{tabular}                                                   & Confidentiality &                     \cite{li2020privacy, javaid2020scalable, shi2019dynamic, lei2017blockchain}                                                                                \\ \hline
Collusion                                                                 & \begin{tabular}[c]{@{}c@{}}A member of the IoV\\ network colludes with \\ an adversary to get \\ them access\end{tabular}                              & \begin{tabular}[c]{@{}c@{}}Authenticity, Confidentiality,\\ Data loss, Physical security\\ of the IoV network resources\end{tabular}   &    \begin{tabular}[c]{@{}c@{}} \cite{zheng2019traceable, li2019toward}                                       \end{tabular}                                                         \\ \hline
Forgery                                                                   & \begin{tabular}[c]{@{}c@{}}Forging the signature \\ of a member of the IoV \\ network\end{tabular}                                                     & \begin{tabular}[c]{@{}c@{}}Authenticity, Physical security\\ of the IoV network resources\end{tabular}                                 &  \begin{tabular}[c]{@{}c@{}}  \cite{liu2019novel, liu2019blockchain, sutrala2020design, tan2019secure, shen2020blockchain}    \end{tabular}                                                                                             \\ \hline
\end{tabular}
\label{tab:network_attack}
\end{table*}

\begin{enumerate}
\item \textbf{Replay attack:}
In this attack, the malicious entity imitates a legitimate user and replays transmitted data in the network that it had previously captured. This attack is carried out to target the authenticity and confidentiality of the system and also to prevent tracking of vehicles in case of a mishap or an accident \cite{mishra2016vanet}.

\item \textbf{Man-in-the-\textcolor{black}{middle} (MITM) attack:} A malicious entity intercepts and changes the messages exchanged between legitimate nodes in the network \cite{ahmad2018man}.

\item \textbf{Impersonation attack:} A malicious entity impersonates another entity in the network and sends messages on its behalf \cite{vangala2020blockchain}. It is also known as a masquerade attack.

\item \textbf{Privileged \textcolor{black}{insider} attack:} An adversary is a privileged user inside the network and can access cryptographic information of the deployed entities in the network, such as certificates and private keys. \cite{vangala2020blockchain}. 

\item \textbf{Spoofing attack:} A malicious entity circulates false data in the network. It may broadcast spurious messages to \textcolor{black}{neighbours} causing traffic accidents and congestion \cite{yang2018blockchain}. It also includes false report generation about network transmission channel availability and other resources \cite{li2019toward}. It can also be in the form of GPS spoofing attacks, where adversaries inject false position data in location-based networks \cite{li2020vehicle}.

\item \textbf{Malicious Third Party (MTP) or bad-mouthing attack:} Most blockchain-based vehicular network schemes use trust management systems where they store the credibility ratings of each vehicle depending on the feedback from other network vehicles about their past \textcolor{black}{behaviour}. Malicious vehicles may generate false ratings and upload them in the blockchain (or other storage places that maintain trust records) to decrease the credibility ratings of genuine vehicles and increase the ratings of the malicious vehicles. This is called MTP, or bad-mouthing attack \cite{yang2018blockchain,su2018secure,kang2019toward}.

\item \textbf{Ephemeral Secret-Key Leakage (ESL) or background analysis attack:} In an ESL or background analysis attack, an adversary breaks the privacy of users by finding out the session keys or true identities of users by analyzing previously used leaked session keys or pseudonymous identities of the users. Adversaries can easily find out the actual identity/\textcolor{black}{behaviour} or break into the communication session if the user continues to use the same pseudonym or secret key every time. Also, in case of secure communication taking place between entities via common sessions, leakage of the session-specific secrets (ephemeral secrets) may endanger the security of the network and further communication sessions. Such attacks can be classified as ESL \cite{vangala2020blockchain} or background analysis attacks \cite{shen2020blockchain}. They can be mitigated by meeting the security requirement of privacy preservation, specifically unlinkability or forward security. Security schemes can use random secrets along with the session-specific secrets to ensure that there is a different session key for every session. A session key of one particular session being revealed will not affect future sessions. This is also called `perfect forward and backward secrecy' \cite{vangala2020blockchain}. Vehicles can be allowed to use a set of randomly generated pseudonyms instead of a single pseudonym to prevent the malicious vehicles from linking their true identity with the pseudonym by background analysis.

\item\textbf{{Sybil attack}:} Sybil attack takes place when a malicious entity tries to control a substantial part of the network by introducing a large number of pseudo-identities or pseudonyms at the same time \cite{douceur2002sybil}. It can use them to illegally alter the reputation values of other nodes (i.e., Sybil attack may lead to bad-mouthing attack). Most vehicular networks offer a feature of forwarding announcements anonymously, which makes them vulnerable to Sybil attacks. Wang et al. \cite{wang2019bsis} discuss how Sybil attacks can take place in an energy trading scenario. In \cite{li2018creditcoin}, the authors mitigate Sybil attacks by using a threshold cryptographic technique based on Lagrange Interpolation, which ensures that only a fixed number of different private keys are used to generate the signature.

\item\textbf{{Denial of Service (DoS) attack}:} In a DoS attack, an adversary attempts to make a network resource or an online service unavailable to legal users by flooding the network with illegitimate service requests \cite{carl2006denial}. Some examples include the HTTP-DoS attack on Iranian sites, which brought down important government sites \cite{chonka2011cloud} and the DoS attack on American Health Care government site \cite{gupta2017taxonomy}. Li et al. \cite{li2019iterative} discuss DoS attacks in the context of electricity trading in vehicular networks where the attacker needs to compromise a huge number of blockchain nodes and control them to send transactions to the target to prevent the target node from participating in the electricity trading.

\item\textbf{{Distributed Denial of Service (DDoS) attack}:} The DDoS attack is similar to a DoS attack, but a DoS attack is executed through a single node whereas, the DDoS attack is carried through multiple nodes in the network. The DDoS attack is generally carried out through `botnets'. Firstly, several nodes are kept under control by infecting them with malware. These compromised nodes are used simultaneously to attack a specific target machine. This network of compromised nodes that can communicate with each other is called a botnet. Li et al. \cite{li2019toward} discuss DDoS attacks in an advertisement dissemination scenario in vehicular networks. Feng et al. \cite{feng2019bpas} proposed a blockchain-assisted authentication mechanism for mitigating DDoS attacks along with protecting against other common attacks such as replay and impersonation attacks.

\item\textbf{{Repudiation attack}:} It refers to the denial of participation by an entity in all or part of communications in vehicular networks. For example, a vehicle driver can deny operating a credit card purchase, or adversaries can abuse anonymous authentication techniques to run away with liabilities \cite{li2014acpn}. This attack scenario can be mitigated by meeting the security requirement of non-repudiation. Li et al. \cite{li2019iterative} discuss mitigation of repudiation attacks with the help of digital signatures in a consortium blockchain-based electric vehicle charging and discharging scheme.

\item\textbf{{Eavesdropping attack}:} The dictionary meaning of eavesdropping means the act of secretly listening to other people talking \cite{bighash2020model}. In a vehicular network scenario, attackers can eavesdrop on the communication between network entities such as vehicles, edge nodes, fog computing nodes \cite{li2020privacy}, and RSUs. In \cite{shi2019dynamic}, the authors present a formal security proof of how their framework encrypts the system using RSA (with 2028-bit encryption algorithm) to resist eavesdropping attacks. Formal security proof such as Burrows–Abadi–Needham (BAN) logic can also be used to determine whether a set of exchanged messages is secured against eavesdropping or not. 

\item \textbf{{Collusion attack}:} It is a type of security attack in which a network node makes a secret agreement with an adversary. The adversary can use the compromised node to carry out malicious activities to exploit the system, such as collecting confidential information, executing sophisticated attacks, injecting false data, etc. \cite{bhuiyan2016collusion}. It can also lead to various kinds of network security attacks and hence can pose a serious threat to the vehicular networks. Some blockchain-based security schemes discuss various types of collusion attacks that can take place in a vehicular network, as given below.
\begin{enumerate}
    \item \textbf{Miner voting collusion:} Collusion between malicious RSUs and compromised stakeholders may lead to the election of false miners who may modify or discard legitimate transaction data during the mining algorithm.
    
    \item \textbf{Block verification collusion:} False results may also be created by the collusion of miners in the block verification stage, which can be a stepping stone to launching a double-spending attack \cite{kang2019toward}.
    
    \item \textbf{Collusion authentication fraud:} Multiple proxy vehicles with selfish intentions in the network may collaborate for executing authentication frauds by attempting to reorganise the network secrets \cite{liu2020blockchain}. It can be mitigated by increasing the number of proxy vehicles and by storing the authentication transaction records. 
\end{enumerate}

\item \textbf{{Forgery attack}:} It refers to an attack where the adversary tries to recreate a digital signature of a message without knowing the respective signer's private key \cite{Bleumer2005}. It can be mitigated by meeting the security requirement of non-repudiation. According to the authors of \cite{liu2019blockchain}, forgery attack is very common in a vehicular network scenario. They use a digital signature where a secret key is generated by a fully trusted authority called Trusted Authority (TA) and stored in a physically isolated Tamper-proof Device (TPD). In \cite{liu2019novel}, the authors use an encrypted signature to prevent adversaries from forging the signature of a vehicle for gaining control of a majority of network resources. Tan et al. \cite{tan2019secure}, and Sutrala et al. \cite{sutrala2020design} formally prove that their frameworks are secure against forgery \textcolor{black}{attacks} using the random oracle model for verification.
\end{enumerate}

Table \ref{tab:network_attack} lists various blockchain-based network-specific attacks, security schemes proposed for vehicular networks and the security attacks mitigated in each of these schemes.

\subsection{Authentication Techniques}
\label{sec:auth_tech}
Authentication is an essential part of any security framework and plays an important role in mitigating most security attacks\cite{alladi2020lightweight,alladi2020harci}. We list some of the popular authentication techniques used in blockchain-based vehicular security frameworks below.

\begin{enumerate}
    \item \textbf{Mutual/Two-way authentication:} In this type of authentication, both the client (vehicle) and the access point (RSU) authenticate each other. The client verifies that it is opening the session with a legitimate access point, and the access point is sure that it is opening a session with an authorized client. Authenticating access points helps in mitigating MITM attacks, and authenticating clients help in mitigating replay attacks \cite{welch2003wireless}. Vangala et al. \cite{vangala2020blockchain} discuss how mutual authentication is established between a cluster head and a vehicle within that cluster (V2CH) and between cluster head and an RSU (CH2RSU) in the authentication phase of their proposed framework. In \cite{huang2018lnsc}, the authors formally prove how their security model achieves mutual authentication between the electric vehicle and the charging pile.
    
    \item \textbf{Anonymous authentication:} It is a type of authentication scheme that preserves the privacy of the vehicles during the authentication process. It is also referred to as privacy-preserving authentication \cite{lu2018survey}. Privacy means that the users have \textcolor{black}{the} full right to control their personal information and reserve the right to choose which data is shared with others. Anonymity refers to the quality of being unidentifiable within a finite set of users. Anonymity is achieved by the help of pseudonyms, which are bit strings used as unique identifiers for authentication without any personally identifiable information. Hence using them helps in authenticating a specific entity without being aware of its real identity, i.e., anonymous authentication. Anonymous authentication schemes can be broadly classified into five categories \cite{lu2018survey}.
    \begin{enumerate}[i.]
        \item Schemes based on symmetric cryptography
        \item Schemes based on Public Key Infrastructure (PKI)
        \item Schemes based on identity-based signature
        \item Schemes based on certificate-less signature
        \item Schemes based on group signature
    \end{enumerate}
    
    Anonymous authentication is commonly employed in vehicular networks for privacy preservation \cite{yao2019bla}. For example, in \cite{li2019toward}, the authors use anonymous authentication for a blockchain-based anonymous fair anonymous ad-dissemination in vehicular networks. Yao et. al.\cite{yao2019bla} proposed a blockchain-assisted lightweight anonymous authentication scheme for distributed vehicular fog services.
    
    \item \textbf{Certificate-based authentication:} In certificate-based authentication, the first entity authenticates itself to the second entity with the help of a digital certificate associated with the first entity \cite{falk2016method}. A digital certificate contains a public key for the corresponding entity, and the owner of the certificate can be confirmed with the help of a digital signature in the certificate. The digital signature is verified by a certificate issuing unit, or more specifically, a Certificate Authority (CA) or a TA in the context of vehicular networks. In \cite{vangala2020blockchain}, the authors use a blockchain-enabled certificate-based authentication scheme (BCAS-VADN) for empowering vehicles to securely report the transactions related to accident detection and notification of their own or \textcolor{black}{neighbouring} vehicles to the cluster head. Zhou et al. use certificate-based authentication in a blockchain and edge computing enabled vehicle-to-grid-energy trading \cite{zhou2019secure}. It has also been used by Kang et al. \cite{kang2017enabling} in their localized peer-to-peer electricity trading model for plug-in hybrid vehicles. They employ a TA such as a government department for implementing certificate-based authentication.
    
    \item \textbf{Batch verification:} In batch verification, instead of a single message/entity, a batch of messages/entities are verified together. According to \cite{zhang2019data}, batch verification can be explained as follows:
        
    Assuming a random selection of an integer $i$ between 1 and $n$ (both included), where $n$ is a positive integer, the received verification parameters can be \textit{Ver\textsubscript{i}(AID\textsubscript{i}, S\textsubscript{i}, M\textsubscript{i}, C\textsubscript{i})}. Here \textit{C\textsubscript{i}} is the randomisation parameter, \textit{M\textsubscript{i}} is the message, \textit{S\textsubscript{i}} is the final signature information and \textit{AID\textsubscript{i}} is the pseudonym generated by the vehicle. The plurality of the authentication parameters generated during batch verification for the messages shared between the network nodes is \textit{BatchVer\textsubscript{n}((AID\textsubscript{1}, S\textsubscript{1}, M\textsubscript{1}, C\textsubscript{1}),...,(AID\textsubscript{n}, S\textsubscript{n}, M\textsubscript{n}, C\textsubscript{n})}. If each of the $n$ signatures is legal then the batch verification is passed \cite{camenisch2007batch}. If any one or more of the $n$ signatures are invalid, the batch verification fails. Batch verification can also be divided into three types:
    \begin{enumerate}[i.]
        \item Verify that a signer signs a different message
        \item Verify that different signers sign the same message
        \item Verify that different signers sign different messages
    \end{enumerate}
    Batch verification can help security schemes cope up with the time delay when verifying multiple messages. In \cite{chen2019secure}, the authors use batch verification in an energy blockchain-based electricity trading framework for EVs. Lin et al. \cite{lin2020bcppa} used a modified ECDSA encryption scheme for implementing batch verification. Also, in \cite{tan2019secure}, the authors formally explain how the batch verification is implemented in the authentication phase of the proposed security framework for blockchain-based VANETs.
    
     \item \textbf{Zero Knowledge Proof (ZKP):} The ZKP algorithm is used when two entities are to be authenticated without the exchange of secret information \cite{singh2020blocked}. The concept was first given by Goldwasser et al. \cite{lee2006distributed}, and since then, it has found many uses in identification and authentication protocols. A prover must demonstrate knowledge of a secret to some verifier through several interactive rounds. Each round consists of a challenge from the verifier and a response from the prover. During the entire process, the prover does not reveal any information whatsoever, to either the verifier or to any third party. In general, it satisfies two basic security properties \cite{li2019toward}: (1) Soundness - It means that the verifier will never accept an invalid result, and (2) Zero-Knowledge - It means that no leakage of any information occurs during the proof. It has been used in many blockchain-based vehicular network frameworks. Singh et al. \cite{singh2020blocked} used it in a blockchain-based data processing framework in edge envisioned V2X environment, and Li et al. \cite{li2019toward} used it in a blockchain-based fair and anonymous ad-dissemination vehicular network.
   
    \item \textbf{Threshold authentication:} It is a standard method to prove the reliability of messages in a vehicular network. In this technique, the receiver node only accepts a message when it is confirmed by a threshold number of vehicles in the network \cite{li2018creditcoin}. Message aggregation in the network is a good way to realize threshold authentication and also to reduce network overhead. Li et al. use a threshold ring signature scheme in \cite{li2018creditcoin} to design their framework, ``CreditCoin'' which is a privacy-preserving \textcolor{black}{blockchain-based} incentive announcement network for communication of smart vehicles. It is also used by Lin et al. \cite{liu2019blockchain} in their blockchain-based trust management announcement scheme for VANETs.
\end{enumerate}

Table \ref{tab:auth_tech} lists the various security frameworks along with the authentication techniques they use.

\begin{table*}[]
\centering
\caption{\textcolor{blue}{Survey of authentication techniques used by blockchain-based security frameworks for vehicular networks.}}
\label{tab:auth_tech}
\begin{tabular}{|c|c|c|c|c|c|c|}
\hline
\rowcolor[HTML]{C0C0C0} 
\textbf{Ref.} & \textbf{Mutual/Two-way} & \textbf{Anonymous} & \textbf{\textcolor{black}{Certificate} based} & \textbf{Batch verifcation} & \textbf{ZKP} & \textbf{Threshold} \\ \hline
\cite{vangala2020blockchain} & \cmark & \cmark & \cmark & - & - & - \\ \hline
\cite{zheng2019traceable} & - & \cmark & - & - & - & - \\ \hline
\cite{zhang2019data} & \cmark & \cmark & - & \cmark & - & - \\ \hline
\cite{huang2018lnsc} & \cmark & - & - & - & - & - \\ \hline
\cite{li2019iterative} & \cmark & \cmark & - & - & - & - \\ \hline
\cite{li2019toward} & - & \cmark & - & - & - & - \\ \hline
\cite{zhou2019secure} & - & - & \cmark & - & - & - \\ \hline
\cite{li2020privacy} & \cmark & - & - & - & - & - \\ \hline
\cite{javaid2020scalable} & \cmark & - & \cmark & - & - & - \\ \hline
\cite{yao2019bla} & - & \cmark & - & - & - & - \\ \hline
\cite{liu2019blockchain} & - & - & - & - & - & \cmark \\ \hline
\cite{sutrala2020design} & \cmark & - & - & \cmark & - & - \\ \hline
\cite{wang2019bsis} & - & - & \cmark & - & - & - \\ \hline
\cite{tan2019secure} & \cmark & - & \cmark & \cmark & - & - \\ \hline
\cite{singh2020blocked} & - & - & - & - & \cmark & - \\ \hline
\cite{lei2017blockchain} & - & - & \cmark & - & - & - \\ \hline
\cite{lin2020bcppa} & - & - & \cmark & \cmark & - & - \\ \hline
\cite{li2018creditcoin} & - & - & - & - & - & \cmark \\ \hline
\cite{li2018efficient} & - & - & \cmark & - & \cmark & - \\ \hline
\cite{li2020consortium} & \cmark & \cmark & \cmark & - & - & - \\ \hline
\cite{wang2020blockchain} & - & - & \cmark & - & - & - \\ \hline
\cite{zhang2019adaptive} & - & - & - & \cmark & - & - \\ \hline
\cite{zhang2020bsfp} & - & \cmark & - & - & - & - \\ \hline
\cite{lu2018privacy} & - & \cmark & - & - & - & - \\ \hline
\cite{dai2020deep} & - & - & \cmark & \cmark & - & - \\ \hline
\cite{baza2019b} & - & - & - & - & \cmark & - \\ \hline
\cite{wang2020consortium} & - & \cmark & - & - & - & - \\ \hline
\cite{wang2020towards} & - & \cmark & - & - & - & - \\ \hline
\end{tabular}
\end{table*}

\subsubsection*{\textbf{Summary}}
In this section, different blockchain-based frameworks are categorised based on security requirements like decentralization, traceability, etc., protection provided from several network-specific security attacks like DoS, spoofing, Sybil, etc., and authentication techniques like mutual, anonymous, etc.

\section{Categorisation based on Blockchain Features}
\label{sec:bc_perspective}
Understanding the different blockchain platforms and types is essential for understanding how blockchains have been used in vehicular network applications. For the sake of clarity, in this section, we categorise the different blockchain-based security works based on the following category types:
\begin{enumerate}[i.]
    \item Blockchain platform
    \item Blockchain type
    \item Consensus algorithm
\end{enumerate}

\begin{figure*}
        \centerline{\includegraphics[width =1.5 \columnwidth]{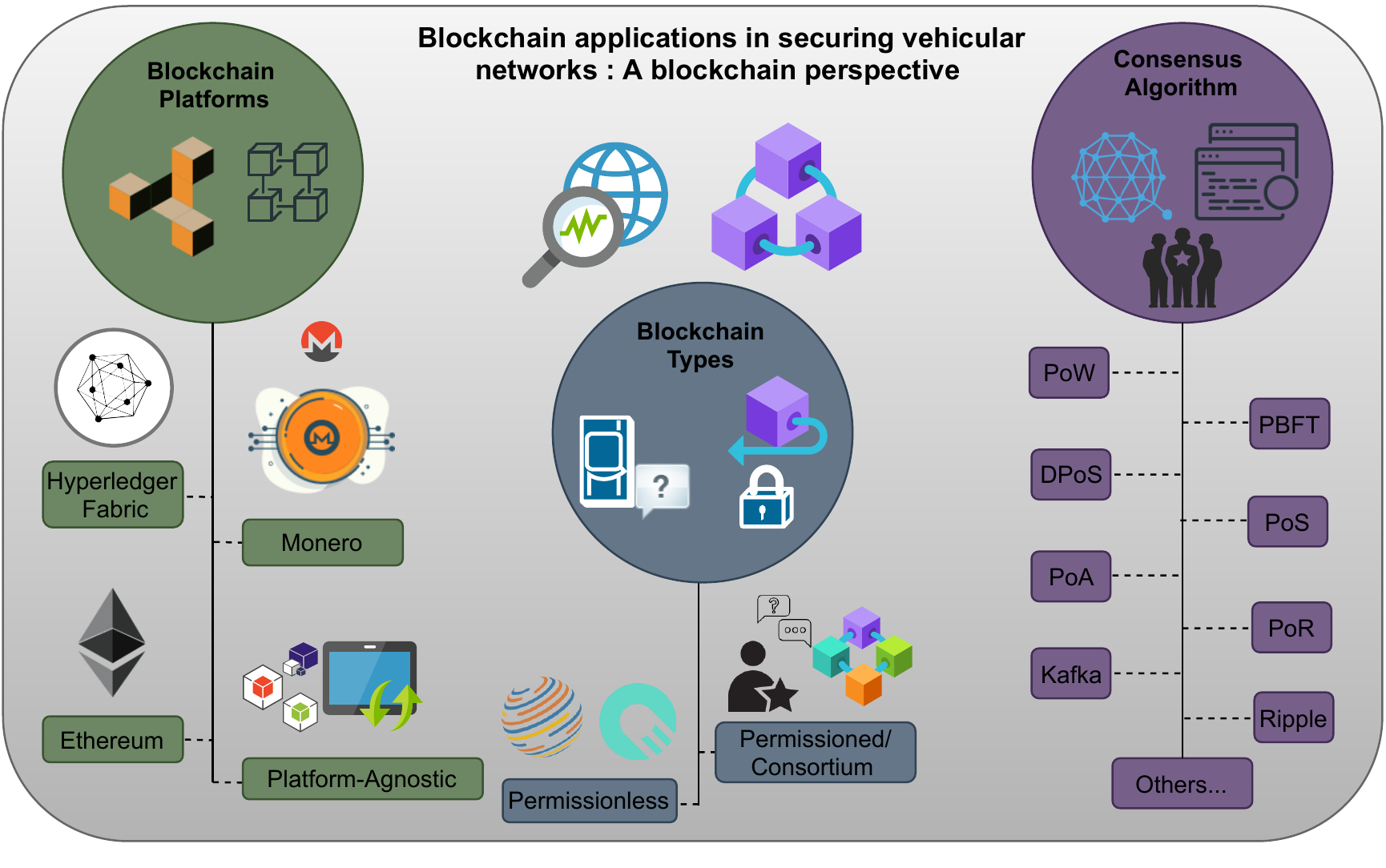}}
        \caption{Overview of the blockchain perspective section.}
        \label{fig:BC_perspective}
\end{figure*}

\subsection{Blockchain Platform}
Since Bitcoin was released as open-source software in 2009, many other platforms have emerged as a result of wider application scenarios of blockchain. These platforms are generally an umbrella service, that comes with their toolkits and ecosystems for developers to build their blockchain applications. The more advanced the platform, the more varied options it offers - for instance, the Hyperledger project offers several different frameworks with different consensus mechanisms and is designed for different applications. Ethereum is another popular platform; while Hyperledger is not strictly associated with any token, Ethereum (like Bitcoin) is associated with its cryptocurrency (Ether/ETH) that has a standard conversion rate with real currency. This is useful, for instance, where blockchain is used as a stand-in for real-world market interactions; in the example of an energy trading application, ETH earned through mining or some other activity can be used directly to purchase energy at charging stations. Apart from cryptocurrency, the parameters by which blockchain platforms generally differ are scalability, throughput, consensus mechanism, crash fault tolerance, Byzantine fault tolerance, smart contract support (and the language used to program them as well) - there is a great amount of variation in many of the technical specifications. The surveyed research works fit into the following categories:

\begin{enumerate}
    \item \textbf{Hyperledger Fabric (HLF):} Hyperledger Fabric, maintained by the Hyperledger community, is an open-source blockchain project. The project provides a few permissioned DLT platforms, designed for use in enterprises \cite{hyperledger_arch_whitepaper}. It is the first DLT platform to support smart contracts authored in general-purpose programming languages such as Java, Go, and Node.js. It is also permissioned, i.e., unlike with a public permissionless network, the participants are known to each other. In \cite{gao2019blockchain}, the authors use HLF (v1.0.2) as their blockchain platform, and in \cite{lu2019blockchain}, an extended HLF architecture that employs a modified blockchain structure, is used to provide a privacy-preserving authentication scheme against potential attacks such as vehicle impersonation and broadcasting forged messages. 

    \item \textbf{Ethereum:} Ethereum is a blockchain platform that allows its users to create smart contracts in Solidity, a Turing-complete language. Smart contract owned accounts function similar to user accounts, except that they contain executable bytecode, which governs the \textcolor{black}{behaviour} of the smart contract account. When a user transacts with the smart contract-owned account, the stored code is executed, and the Ethereum Virtual Machine records the change. It has been used in many frameworks, for example, in \cite{zhang2019adaptive}, the authors use an Ethereum blockchain in their adaptive traffic control mechanism, and in \cite{liu2019electric}, an Ethereum private chain is used for power-sharing between EVs on a V2G network.

    \item \textbf{Monero:} Monero is a cryptocurrency released in 2014 which primarily focuses on privacy and is an open-source protocol based on CryptoNote. A concept called obfuscated public ledger used in Monero ensures that anyone can broadcast or send transactions, but no observer from outside can find the destination, source, and amount \cite{van2013cryptonote,monero}.  It is used in Blockchain-Based Anonymous Rewarding Scheme (BBARS) for V2G networks, proposed by Wang et al. in \cite{wang2019bbars}. Vehicles can sell stored energy back to the grid and make a profit. The BBARS system model involves a central aggregator that authorizes and pays vehicles through an intermediate local aggregator for supplying excess power to the grid, but without either the payer or payee requiring each others’ addresses or identities to complete the transaction. Elliptic Curve Cryptography is used to ensure security for all transactions.

    \item \textbf{Platform-agnostic:}
    Many proposed research models only utilize the essential features of blockchain, for instance, the decentralized peer-to-peer sharing mechanism and immutability. These, along with a few other properties, are common to every blockchain - they are, in some sense, properties without which software cannot be called `blockchain' in the first place. So it is entirely possible to abstract away the details of the blockchain and design a system that is flexible enough to be used with any blockchain platform. There have been models proposing game theory-based security \cite{kim2019enhanced}, malicious content detection \cite{ahmad2020marine}, traffic signal management \cite{cheng2019sctsc}, energy saving by distributed clustering \cite{sharma2018energy}, GPS accuracy enhancement \cite{song2020blockchain}, platooning \cite{chen2019smart}, data sharing \cite{kang2019toward}, and authentication \cite{wang2019improved} that do not use any properties of a blockchain that are tied to a particular platform or framework, and are therefore equally applicable to all blockchain-based vehicular networks.
\end{enumerate}

\subsection{Blockchain Type}
There are primarily two types of blockchain networks - permissioned and permissionless. Permissioned blockchains reserve the right to validate and add blocks to a select few nodes, while in permissionless blockchains, any node that joins is also permitted (or expected) to validate and add blocks to the blockchain. This is heavily application-dependent. For instance, in the case of energy trading, one would want only the charging stations that belong to a registered company to provide a platform for trading. Therefore, the ability to add transactions must be restricted to nodes that are charging stations, and not the vehicles themselves. A permissioned blockchain would be appropriate for this. Permissioned blockchains also have higher throughput, because a significant portion of the latency is due to the peer-to-peer interactions - if that is minimized, then the network can handle more transactions per second. It must be noted that some aspect of decentralization is sacrificed in a permissioned blockchain - there is a trade-off that needs to be made between throughput and level of decentralization. There are also some applications like crowdsourcing and data sharing, that demand a permissionless blockchain. Some representative research works out of the surveyed set have been selected to illustrate these ideas in more detail below.

\begin{enumerate}
    \item \textbf{Permissioned}: Zheng et al. \cite{zheng2019traceable} proposed a blockchain system for VANETs that provides conditional privacy - here, a certifying authority is the only entity that may validate blocks. This is an authentication system where proof of legitimate identity is stored on the blockchain - allowing any vehicle to add a block to it defeats the purpose, and therefore this must be a permissioned blockchain. BloCkEd is another proposed research work that enables effective edge computing in vehicular networks - using the idea of migrating containers between edge devices using blockchain, tasks from vehicles can easily be offloaded to the edge computing devices and results returned no matter what path the vehicle takes \cite{singh2020blocked}. This system necessarily requires only edge computing devices to have the authority to validate blocks, since vehicles need to be on the same network to interface with the edge devices but cannot participate in migrating the containers. Su et al. propose a framework for energy sharing \cite{su2018secure} in which only the charging piles are given the authority to validate transactions. Apart from improving throughput, this ensures that all transactions happen only via the trusted smart grid operator.
    
    \item \textbf{Permissionless}: This type of blockchain is not as popular, due to scalability issues. However, some crowdsensing applications \cite{zhang2020decentralized, lai2019spir, wang2020blockchain} deploy a permissionless blockchain because the transactions of the blockchain are pieces of information that indicate something about the condition of the road. The validating nodes must be the peer nodes that also witness the same condition, not some trusted authority in this case - like how a permissionless blockchain defeats the purpose of authentication, in this case, a permissioned blockchain defeats the purpose of crowdsensing. Javaid et al. \cite{javaid2020scalable} propose a permissionless blockchain for trust management - the problem of scalability is managed by a dynamic proof-of-work consensus where the difficulty threshold is constantly changed based on the required security fidelity, scale, and throughput.
\end{enumerate}

\subsection{Consensus Mechanisms}
There are many different reasons for choosing one consensus mechanism over the other. PoW is by far the most common consensus algorithm, but it is inefficient and wastes energy - a more economical alternative is PoS. Neither of these provides Byzantine Fault Tolerance, which is the quality of being resistant to the famous Byzantine Generals Problem vulnerability. What follows is an illustration of the different types of research works classified based on the consensus mechanism used.

\begin{table}[h]
\centering
  \caption{Categorisation based on consensus algorithms \textcolor{black}{used in vehicular network security papers.}}
    \begin{tabular}{p{0.18\textwidth}p{0.25\textwidth}}
    \hline
    \hline
    Consensus & Ref. \\
    \hline
    \hline
    Proof of Work (PoW)   & \cite{zheng2019traceable, javaid2020scalable, liu2019blockchain, shrestha2019regional, zhou2019blockchain, chen2019secure, liu2020blockchain, lei2017blockchain, yang2020ldv, zhang2020decentralized, zhou2019secure, liu2019novel, yang2018blockchain, lu2018privacy, lin2020blockchain, huang2020securing, liao2020blockchain} \\
    \hline
    Practical Byzantine Fault Tolerance (PBFT)  & \cite{vangala2020blockchain, zhang2019data, sheikh2019secured, gao2019blockchain, yao2019bla, wang2019improved, liu2019blockchain, feng2019bpas, fu2020autonomous, luo2019blockchain, chen2019toward, lin2020blockchain, yao2019lightweight} \\
    \hline
    Delegated Proof of Stake (DPoS)  & \cite{song2020blockchain, kang2019toward, zhang2019blockchain, fu2020autonomous, lu2020blockchain, su2020lvbs} \\
    \hline
    Proof of Stake (PoS) & \cite{yang2018blockchain, li2018efficient, deng2020electronic, kang2018blockchain, wang2020consortium} \\
    \hline
    Proof of Activity (PoA)  & \cite{li2019toward, xia2020bayesian, liu2019electric, lin2020bcppa} \\
    \hline
    Proof of Reputation (PoR)  & \cite{chai2019proof, wang2019bsis, liu2020blockchain} \\
    \hline
    KafKa & \cite{li2019iterative, li2020consortium} \\
    \hline
    Ripple consensus & \cite{wang2019improved, li2018creditcoin} \\
    \hline
    Attribute based consensus & \cite{cheng2019sctsc} \\
    \hline
    Proof of Online Duration & \cite{li2020privacy} \\
    \hline
    Delegated Byzantine Fault Tolerance (DBFT)  & \cite{su2018secure} \\
    \hline
    Redundant Byzantine Fault Tolerance (RBFT) & \cite{zhang2019blockchain} \\
    \hline
    Zero Knowledge Proof (ZKP)   & \cite{singh2020blocked} \\
    \hline
    Proof of Knowledge (PoK) & \cite{chai2020hierarchical} \\
    \hline
    Proof of Utility (PoU) & \cite{dai2020deep} \\
    \hline
    DPOSP (PBFT-DPoS hybrid) & \cite{sun2020blockchain} \\
    \hline
    Proof of Elapsed Time (PoET) & \cite{iqbal2020blockchain} \\
    \hline
    Proof of Event (PoE) & \cite{yang2019blockchain} \\
    \hline
    AlgoRand & \cite{zhang2020decentralized} \\
    \hline
    \hline
    \end{tabular}
  \label{tab:addlabel}
\end{table}

\begin{enumerate}
    \item \textbf{Proof-of-Work (PoW):}
    Zheng et al. propose a model for pseudonymous message exchange among vehicles. The consensus adopted in the blockchain composed of all RSUs is PoW. This algorithm requires guessing a random number by trial and error, which is time-consuming and costly but is easy to verify. As long as the number of malicious nodes does not exceed half of all the RSUs, the transaction information cannot be modified, and the longer the blockchain, the better will be its security \cite{zheng2019traceable}. Micro-Blockchain-Based Dynamic Intrusion Detection (MBID) is a mechanism proposed by Liang et al. \cite{liang2019mbid} for intrusion detection. Micro-blockchains are nested into a macro-blockchain, and together they provide strategies for detecting intrusions into the vehicular network. The main tasks of micro-blockchains are to store data and intrusion samples as well as provide intrusion detection strategies if they are available in the micro-blockchain. Macro-blockchains store all the models and the collected intrusion intelligence, and provide intrusion detection strategies to vehicles when those strategies are not found in the appropriate micro-blockchain. Network slicing is used to deploy micro-blockchains in the same region. PoW is used as the consensus algorithm for the macro-blockchain.

    \item \textbf{Proof-of-Reputation (PoR):}
    A blockchain architecture that employs PoR is proposed for resource sharing in IoV networks. The lightweight blockchain handles trust management and privacy preservation. A reputation value is assigned to indicate the degree of trust of a vehicle using which consensus is established. An RSU can publish a block only if it has collected the highest reputation sum for transactions \cite{chai2019proof}. Chai et al. \cite{chai2019proof} propose a blockchain model for resource sharing that uses PoR to generate blocks. Reputation is decided based on an accumulation of historical records, with a saturating upper limit on reputation to prevent any node from monopolizing the blockchain. The node with the highest reputation creates blocks of a fixed number of transactions. Higher reputed nodes are more likely to have the blocks they validate accepted into the main blockchain, as well as pay a smaller cost for purchasing computational resources.

    \item \textbf{Proof-of-Stake (PoS):}
    Delegated proof of stake (DPoS) is used for achieving network consensus in some works. Each RSU in the network, based on its stake, selects witnesses to participate in the blockchain system. The top few witnesses (RSUs) with the highest vote count win the verification rights. The elected witnesses create new blocks in sequence as assigned and get some rewards. It accelerates the speed of block creation and transaction verification because of the reduction of verification nodes \cite{song2020blockchain}. A coalition of a certain number of RSUs above a threshold is required for vehicular authentication in a consortium blockchain-based Vehicular Social Networks (VSNs). If the stake is replaced by a rank of activity (which indicates the RSU’s message relay participation rate in the leader selection process), this modified PoS is highly efficient \cite{shen2020blockchain}. The authors in \cite{kang2019toward} propose an enhanced DPoS consensus scheme with a two-stage security solution for secure vehicle data sharing in blockchain-based IoV. In the first stage, miners are selected by voting based on reputation using a subjective logic method. In the second stage, standby miners are incentivized to participate in block verification.

    \item \textbf{Byzantine Fault Tolerance (BFT):}
    Byzantine Fault Tolerance is simply resistance to the Byzantine Generals Problem, a situation where a certain number of peer nodes may not be able to reach an agreement, and it is also unclear to the honest nodes which ones are unable to reach an agreement. Even if some of the nodes fail or behave maliciously, the system must function regardless of whether the other nodes know of those events or not. In an energy market, malicious operators will heavily threaten EVs ’ security and privacy. A contract-based secure charging scheme for EVs with a permissioned blockchain system is proposed to satisfy EVs’ individual needs for energy sources while maximizing the operator’s utility. Delegated Byzantine Fault Tolerance (DBFT) consensus algorithm based on reputation efficiently achieves consensus. DBFT provides $f = (K - 1)/3$ fault tolerance to a consensus system which comprises $K$ consensus nodes \cite{su2018secure}. A scheme is proposed where vehicular public key infrastructure (VPKI) is used with a permissioned blockchain and a fragmented ledger to provide comprehensive forensic services for accident investigations. A leader is selected randomly for every block created from among the validator nodes (e.g., RSUs). The leader proposes a block to the network, and the validators run Byzantine protocols (e.g., PBFT) to reach a consensus on this block. These protocols are resilient to malicious actions of the leader and the participants \cite{cebe2018block4forensic}. A blockchain-based secure data sharing system is presented, which consists of a parent chain and auxiliary chains. A consensus mechanism, such as PBFT, is used to ensure message consistency. Here, the announcement message is signed by every vehicle that witnessed the event - using a short-term secret key based on a multi-signature scheme. At a particular threshold of signers, the announcement message is deemed trustworthy \cite{zhang2019blockchain}.

    \item \textbf{Other mechanisms:}
    A hybrid type of blockchain (PermiDAG) is developed that consists of a local Directed Acyclic Graph (DAG) and permissioned blockchain. It is made up of the local DAG and is also partition-tolerant, which means the blockchain can run effectively even with a partial network. Also by integrating different learning parameters into the network \textcolor{black}{can} lead to the features of learned models being further verified through a two-stage verification \cite{lu2020blockchain}. Attribute-based blockchains can be used to improve traffic management in the IoV. \cite{cheng2019sctsc} proposed a model that uses this concept. A node in the attribute-based blockchain is distinguished by its anonymous identity (pseudonym) and by a set of fuzzy identities (locations, the direction of travel) - both of which are attributes. Simple nodes may read and generate new messages while consensus nodes cannot write/generate messages in a blockchain directly. But it is the consensus nodes that run the consensus algorithm to attain data consistency. To improve the urban traffic condition and reduce accidents, a platoon-driving model is suggested by \cite{chen2019smart} for autonomous vehicles. Vehicles travel in platoons, with more experienced and credible vehicles serving as the platoon heads. Credit is a collective confidence measure accrued over a while to resolve whether a vehicle is trustworthy or not. In a platoon, the reputation value increases the likelihood that the vehicle in question will become the next platoon head. This ensures that the position of the platoon head is not monopolized.
\end{enumerate}

\begin{figure}[]
        \centerline{\includegraphics[width = 1\columnwidth]{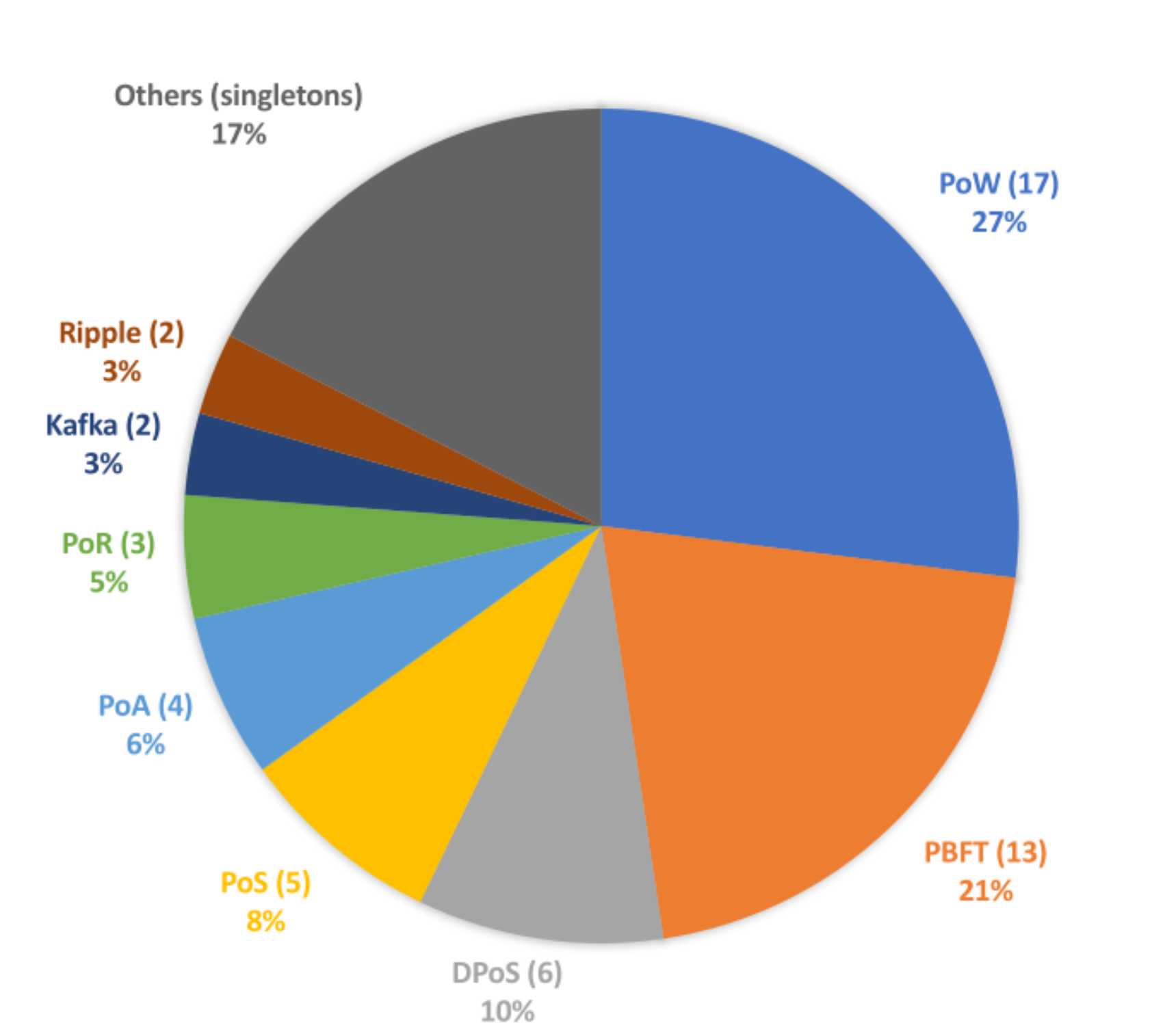}}
        \caption{Pie chart showing the usage of various consensus mechanisms in the surveyed works.}
        \label{fig:pie_consensus}
\end{figure}

Fig. \ref{fig:pie_consensus} presents a pie chart showing the usage of various consensus mechanisms in the surveyed works.

\subsubsection*{\textbf{Summary}}
This section discusses the categorisation of a few works from the perspective of blockchain features they provide. The first categorisation was done based on blockchain platforms like Ethereum, HLF, etc. Secondly, different approaches were grouped based on the type of blockchain, i.e. permissioned and permissionless. Moreover, we categorise blockchain-based frameworks used for IoV based on the consensus mechanism employed by them.

\section{\textcolor{blue}{Compilation of Simulation Tools used in State-of-the-Art BIoV Frameworks}}
\label{sec:simulation}
In this section, we present a compilation of popular simulation tools used for simulating both blockchain and vehicular networks.

\subsection{Tools for Simulating Blockchain}
MIRACL (Multiprecision Integer and Rational Arithmetic Cryptographic Library) is a popular open-source C library that is widely regarded as the gold standard for elliptic curve cryptography (ECC). Blockchain uses several cryptographic primitives, including hashes and asymmetric key encryption, and can therefore be programmed with this library. The authors in \cite{zhang2019blockchain, wang2019bbars} use this library for simulations. PBC (Pairing-based Cryptography) is another library that simulated only mathematical operations of pairing-based cryptosystems; that is, public and private key system simulations without certifying authorities or signatures. Many of these libraries work in conjunction with the GNU Multiprecision Library (GMP) that allows developers to perform operations on arbitrarily precise numbers, which is useful for cryptography applications and research. Ethereum based blockchain simulations can use Solidity, which is a Turing-complete (i.e. computationally universal) language that can be used to write smart contracts that can carry out virtually any application. The authors in \cite{sharma2018energy, jiang2018blockchain, yin2019efficient} use MATLAB in their blockchain simulations. IBM's Hyperledger Fabric is an open-source blockchain infrastructure that can be used to build blockchain applications in Java and Go. The platform is used for simulating permissioned platforms, with a configurable consensus. Hyperledger Fabric has been used in \cite{lu2019blockchain} for simulation. Hyperledger Caliper is a popular benchmarking tool, that measures throughput, latency, and resource consumption for different blockchain solutions; it can also be used with the Ethereum Platform. Besides these, Python is also used extensively.

\subsection{Tools for Simulating Vehicular Networks}
The authors in \cite{labrador2019implementing} use SUMO and OMNeT++ libraries to simulate the behaviour of vehicular networks in the IoV for blockchain. SUMO is used to simulate traffic, and OMNeT++ is used to simulate network behaviour. Veins \cite{sommer2011bidirectionally} is a simulation tool that couples both these technologies, along with the MiXiM framework that simulates physical layer properties (like radio wave propagation, interference estimation, and signal power consumption). Authors of \cite{wang2019improved} use Veins to simulate the network aspect of their models. Other publications using Veins can be found in \cite{haidari2019veins}. Go is another open-source programming language that is an efficient choice for networking, used in \cite{zheng2019traceable}. TraNS \cite{piorkowski2008trans} is a popular open-source platform that couples the SUMO simulator and NS2. However, it has stopped being actively supported since 2008, and iTETRIS \cite{kumar2010itetris} is the logical extension of TraNS, coupling SUMO and NS3. Node.js and web3.js can also be used to connect and interact with different nodes in the system. Mbed TLS (previously PolarSSL) has been used by the authors of \cite{chen2020blockchain} as a lightweight TLS implementation.

\subsubsection*{\textbf{Summary}}
In this section, we discuss the simulation tools for blockchain networks, like MIRACL, PBC, etc., and the simulation tools used for the vehicular networks, like OMNet++, MiXiM, etc.

\begin{table*}[]
\centering
\caption{Role of other state-of-the-art technologies in blockchain based vehicular networks}
\label{tab:other_tech}
\resizebox{\textwidth}{!}{%
\begin{tabular}{|c|c|l|}
\hline
\rowcolor[HTML]{C0C0C0} 
\textbf{Technology}& \textbf{Ref.}& \textbf{Role of other technology in the network}\\ \hline
 & \cite{vangala2020blockchain} & Cloud server was used as part of the blockchain center for creating, verifying and adding the blocks. \\ \cline{2-3}
& \cite{wang2019improved} & Used trusted cloud servers as part of their system architecture to enhance network operations  
\\ \cline{2-3} 
& \cite{lei2017blockchain} & Some of the security managers are cloud based     \\ \cline{2-3} 
& \cite{li2018creditcoin}      & Used a cloud application server to store and exchange some non-cryptographic information in the network                    \\ \cline{2-3} 
& \cite{li2018efficient}       & Used a cloud server to store the encrypted carpooling data                                                               \\ \cline{2-3} 
\multirow{-6}{*}{\textbf{Cloud Computing}} & \cite{yao2019lightweight}    & A distributed and large cloud system is used for storing the Personally Identifiable Information (PII) of the vehicles. \\ \hline
& \cite{kang2017privacy}       & \begin{tabular}[c]{@{}l@{}} Proposed a pseudonym based privacy preserving scheme for maintaining location privacy specifically in fog \\ computing based IoV networks\end{tabular}                                                     \\ \cline{2-3} 
& \cite{li2020privacy}         & \begin{tabular}[c]{@{}l@{}}Used fog computing to overcome drawbacks of a cloud-based centralized management and to provide local \\ computing capabilities with low latency\end{tabular}
\\ \cline{2-3} 
& \cite{li2018pros}& \begin{tabular}[c]{@{}l@{}} Proposed a route sharing service framework with privacy-preservation where fog nodes pre-process the data and provide \\ anonymous authentication.\end{tabular}
\\ \cline{2-3} 
& \cite{sun2020blockchain}     & Proposed a fog computing based V2V energy trading architecture in social hotspots.
\\ \cline{2-3} 
& \cite{li2018efficient} & \begin{tabular}[c]{@{}l@{}} Proposed a carpooling scheme using vehicular fog computing nodes in a blockchain network \\ that supports conditional privacy, destination matching of users, and auditability of data.\end{tabular} \\ \cline{2-3} 
& \cite{gao2019blockchain}     & Fog computing avoids frequent handovers in the network.\\ \cline{2-3} 
& \cite{yao2019bla}            & \begin{tabular}[c]{@{}l@{}}Proposed a blockchain-based anonymous lightweight authentication mechanism for distributed \\ Vehicular Fog Services (VFS)\end{tabular}
\\ \cline{2-3} 
\multirow{-8}{*}{ \textbf{Fog computing}}              & \cite{iqbal2020blockchain}   & \begin{tabular}[c]{@{}l@{}} Proposed a fog computing based secure framework where RSUs for offload tasks to nearby fog \\ vehicles on the basis of reputation scores which are stored in the blockchain \end{tabular}
\\ \hline
& \cite{li2020vehicle} & Used edge server for running DNN based prediction algorithms for evaluating the positioning error
\\ \cline{2-3} 
& \cite{tan2019secure} & Edge computing infrastructure offers extra computing and storage for vehicles in the network.
\\ \cline{2-3} 
& \cite{zhou2019secure}& Developed a task offloading mechanism based on edge computing to increase the success probability of block creation.
\\ \cline{2-3} 
& \cite{liu2020blockchain}     & \begin{tabular}[c]{@{}l@{}}Proposed a system model based on Dynamic Proxy Edge Computing mode in vehicular networks \\ where the vehicles with edge computing and communication capabilities are the proxy vehicles.\end{tabular}         
\\ \cline{2-3} 
& \cite{lin2020blockchain} & \begin{tabular}[c]{@{}l@{}} Incorporates Mobile Edge Computing (MEC) nodes for providing computing resources and to act as managers of local \\ P2P trading system.\end{tabular}
\\ \cline{2-3} 
& \cite{vangala2020blockchain} & Edge servers analyse transactions and contribute to partial block creation
\\ \cline{2-3} 
\multirow{-7}{*}{\textbf{Edge  computing}}
& \cite{dai2020deep}           & Vehicular edge computing network
\\ \hline
& \cite{ahmad2019realization}  & NDN solves issues posed by IP-based networks
\\ \cline{2-3} 
& \cite{rawat2020blockchain}   & Amalgamation of blokchain and NDN helps in achieving integrity and accountability in the network.
\\ \cline{2-3} 
\multirow{-3}{*}{ \textbf{\begin{tabular}[c]{@{}l@{}}Named Data \\ Networking (NDN)\end{tabular}}}     & \cite{chen2019secure}        & \begin{tabular}[c]{@{}l@{}} NDN facilitates content centric data sharing for IoV. Bottom layer nodes request for service via announcements \\ in the NDN paradigm\end{tabular}
\\ \hline
& \cite{zhang2019blockchain}   & Framework is based on an area control plane, programmable by SDN mechanisms.
\\ \cline{2-3} 
\multirow{-2}{*}{\textbf{\begin{tabular}[c]{@{}l@{}}Software Defined \\ Networking (SDN)\end{tabular}}}                       & \cite{gao2019blockchain}     & SDN guarantees that the control processes are adequately accomplished in the vehicular network                                                                                                                                                            \\ \hline
& \cite{lu2020blockchain}   & Used for efficient data sharing in IoV and quality verification based on reputation in IoV \\ \cline{2-3} 
& \cite{hassija2020dagiov}     & DAG data structure stores and validates transactions in the network\\ \cline{2-3} 
& \cite{gong2020secured}       & Approach based on DAG improves the security and authenticity of the proposed energy management scheme                                                                                                                                                \\ \cline{2-3} 
\multirow{-4}{*}{\textbf{\begin{tabular}[c]{@{}l@{}}Directed Acyclic \\ Graphs (DAG)\end{tabular}}}   & \cite{yang2020ldv}           & DAG based blockchains have high throughput\\ \hline

& \cite{nadeem2019securing} & Use Cognitive Radio technology to solve the problem of spectrum storage in vehicular systems. \\ \cline{2-3} 
\multirow{-2}{*}{\textbf{\begin{tabular}[c]{@{}l@{}}Cognitive Radio \\ Technology\end{tabular}}}                              & \cite{rathee2020crt}         & \begin{tabular}[c]{@{}l@{}}Framework provides security to IoV during spectrum sensing and data transmission with the help of a \\ Cognitive Radio Network\end{tabular}\\ \hline
%\rowcolor[HTML]{EFEFEF}
& \cite{li2020vehicle}         & Used DNN-based prediction algorithm for evaluating the positioning error                                                                                                                                                                           \\ \cline{2-3} 
%\rowcolor[HTML]{EFEFEF} 
\multirow{-2}{*}{\textbf{\begin{tabular}[c]{@{}l@{}}AI - Deep Neural \\ networks (DNN)\end{tabular}}} & \cite{song2020blockchain}    & DNN algorithm used for accurate vehicle positioning   \\ \hline

& \cite{fu2020autonomous}      & Used DRL for modelling the lane changing problem of autonomous vehicles.      \\ \cline{2-3} 
& \cite{lu2020blockchain}   & Used an asynchronous federated learning scheme. DRL is used for selection of nodes.                                             \\ \cline{2-3} 
& \cite{chai2019proof} & Used a DRL based smart contract scheme for pricing based on supply-demand during the resource sharing process. \\ \cline{2-3} 
\multirow{-4}{*}{\textbf{\begin{tabular}[c]{@{}l@{}}AI - Deep \\ Reinforcement\\ Learning (DRL)\end{tabular}}}                & \cite{dai2020deep}           & DRL integrated into a permissioned blockchain for secure and intelligent content sharing in the vehicular network.\\ \hline
%\rowcolor[HTML]{EFEFEF} 
\textbf{\begin{tabular}[c]{@{}l@{}}AI - Dueling Deep \\ Q-Learning\\ (DDQL) approach\end{tabular}}                           & \cite{zhang2019blockchain}   & \begin{tabular}[c]{@{}l@{}}A novel DDQL with prioritized replay approach used to solve optimization problems in vehicular networks \\ (which is modelled as a Markov decision process).\end{tabular}                                                         \\ \hline
& \cite{fu2020autonomous}  & Used DRL for modelling the lane changing problem of autonomous vehicles. \\ \cline{2-3} 
\multirow{-2}{*}{\textbf{\begin{tabular}[c]{@{}l@{}}AI - Federated \\ Learning\end{tabular}}}                                 & \cite{chai2020hierarchical}  & A multilevel federated learning algorithm is used to satisfy the distributed pattern and privacy requirements of IoVs.
\\ \hline
%\rowcolor[HTML]{EFEFEF} 
\textbf{\textcolor{black}{Differential Privacy}} & \textcolor{black}{\cite{hassan2020differential}}  & \textcolor{black}{Integrated blockchain with differential privacy to alleviate privacy concerns in intelligent transportation systems.}\\ \hline
\textbf{5G}                                                                                                                   & \cite{gao2019blockchain}     & \begin{tabular}[c]{@{}l@{}}Analyzes the SDN based blockchain network for effective operation of the VANET systems, especially in 5G and fog\\ computing paradigms.\end{tabular}                                                                                  \\ \hline
\textbf{UAVs}                                                                                                                 & \cite{su2020lvbs}            & \begin{tabular}[c]{@{}l@{}}UAV assisted aerial to ground framework providing secure and efficient data transmission in areas affected by a\\ natural disaster.  \end{tabular}         \\ \hline
%\rowcolor[HTML]{EFEFEF} 
\textbf{PUF}                                                                                                                  & \cite{javaid2020scalable}    & PUFs provide a unique fingerprint to identify each vehicle and build a trust management system.                                                           \\ \hline
\textbf{\begin{tabular}[c]{@{}l@{}}Wireless Link \\ Fingerprints\end{tabular}}                                                & \cite{kamal2020blockchain}   & Used wireless finger prints for encryption.                                 \\ \hline
\end{tabular}
}
\end{table*}

\section{\textcolor{black}{Role of Important Technologies Frequently Used in BIoV Frameworks}}
\label{sec:role_of_other_tech}
In this section, we discuss how various state-of-the-art technologies have been used in securing various vehicular networks such as IoV and VANETs. A table summarizing the role of these technologies in the surveyed works is presented in Table \ref{tab:other_tech}.

\subsection{Cloud and Fog Computing}
\textcolor{blue}{Blockchain based IoVs are systems that require enormous amounts of computation and hardware resources, and technologies that manage the distribution of this load -- like cloud computing -- will eventually play a central role.}

Cloud computing is a notion that hardware and software services can be purchased by those who need it from those who have enough under-utilized infrastructure. Over the Internet, a user may purchase the amount of processing or services they require, and the task will be offloaded to a remote computing unit, seemingly to a ``cloud server". This decoupling of hardware and software allows innovators to come up with applications without having to worry about the required hardware infrastructure. This technology is especially useful for vehicles, where consumers have just as many resources as they have the budget for. Fog computing is an extension of cloud computing, where cloud services are brought closer to the sensors and the embedded systems. In the context of IoV, RSUs can serve as fog units or local gateways to the cloud. \textcolor{blue}{Broadly, cloud computing technology can be used with blockchain-based vehicular networks in two ways}:

\begin{enumerate}
    \item Vehicular clouds (also called Autonomous Vehicular Clouds) \cite{eltoweissy2010towards}, where the nodes are the service providers. Vehicles with excess computational resources can take on computational tasks of other vehicles.  
    \item VANETs that use cloud services, where the RSUs act as gateways to the internet. By providing fog layers, this technology can be made even more effective for vehicular networks \cite{yu2018deployment, lu2019cognitive, wang2019auto}.
\end{enumerate}

\subsection{Edge Computing}
Edge computing is an extension of cloud computing that brings cloud services closer to the edge, where the sensors gather data. This is advantageous in terms of processing speed and decentralization since data does not need to be transmitted over long distances. With the large amounts of data being generated in an IoV network, transmitting that quantity of data over long distances, processing it, and returning the results with low latency is also infeasible. \textcolor{blue}{Consensus algorithms that are run on blockchains are also computationally heavy, and solutions that leverage edge computing in their blockchain architectures can ease the computational load of vehicles in the IoV. Much of the processing work can be relegated to the edge nodes or RSUs in the case of vehicular networks.} Mobile Edge Computing (MEC) architecture \cite{mach2017mobile} has been gaining some attention due to this. The robustness of the blockchain network is also improved since RSUs can also be miners. To ensure optimal quality-of-service (QoS), a suitable trade-off between performance and privacy requirements must be made. The authors of \cite{kang2017privacy} propose a pseudonym-based privacy-preserving scheme for maintaining location privacy specifically in fog computing-based IoV networks. Route-sharing is another important aspect of IoV networks - users can upload their route information in real-time for a variety of services, including platooning and traffic forecasting. The authors of \cite{li2018pros} propose a privacy-preserving route sharing service model where fog nodes pre-process the data and provide anonymous authentication.

\subsection{CCN/NDN/ICN}
Content-Centric Networking (CCN), -- also known as Named Data Networking (NDN) or Information-Centric Networking (ICN) -- is a data communication paradigm that places more emphasis on the content rather than the connection \cite{jacobson2009networking}. It replaces Internet Protocol's end-to-end connection model with a request-reply model \cite{wang2012data}. \textcolor{blue}{Unlike TCP/IP, the addresses of the nodes are not required for data exchange. This approach leads to efficient record broadcast as well as ledger synchronization in blockchain networks. There is no concept of a full node and a light node, which removes the security vulnerability in which the light node is fully dependent on the full node to retrieve and return data \cite{asaf2020blockchain}.} Simulations have proven that this paradigm significantly enhances network performance \cite{mau2014vehicular}. Yu et al. \cite{yu2012content} proposed an NDN routing protocol called Bloom-filter routing, based on bloom filters (which are probabilistic data structures that efficiently convey one of two things about a queried element - either it may be part of the set, or it is not). There is a lot of recent interest in this technology since it is especially applicable for IoV. They have been used in studies like \cite{ahmad2019realization,rawat2020blockchain,chen2019secure} for achieving high integrity and accountability in a vehicular network.

\subsection{Software Defined Networking (SDN)}
SDN is a relatively recent technique that helps in efficient network management in communication networks. Traditional data networks face many issues in extension and updates because of their complex nature in which the data and control planes are coupled together. SDN is a programmable network that reduces network complexity by decoupling the control and data planes. It removes the control functionality from the nodes and transforms them into a simple packet/data forwarding node. \textcolor{blue}{This has many benefits for a network as volatile as a VANET. However, the decoupling of control and data planes also expands the attack surface, and blockchain is a viable solution to addressing that problem. Xue et al. \cite{xue2019research} have conducted an in-depth study of how blockchain may be used to address existing problems with SDN networks.} An external entity called the SDN controller is responsible for controlling the network affairs \cite{gao2019blockchain}. Few studies have used SDN along with blockchain for making their frameworks more robust and efficient. For example, in \cite{zhang2019blockchain}, the authors propose a novel blockchain-based distributed software-defined VANET framework (Block-SDV) based on a programmable area control plane originated from SDN. Also, in \cite{gao2019blockchain}, Gao et al. designed a blockchain-based, SDN enabled IoV environment for fog computing and 5G networks that used SDN to guarantee that the control processes are adequately accomplished in the vehicular networks. Chamola et al. \cite{chamola2020optimal} propose a task offloading framework for cloudlets using SDN to manage resource sharing.

% \begin{figure}
%         \centerline{\includegraphics[width =1 \columnwidth]{Figures/8_normal_vs_DAG.pdf}}
%         \caption{Conceptual comparison of conventional (top) and DAG (bottom) based blockchain structures \cite{yang2020ldv}. }
%         \label{fig:normal_vs_DAG}
% \end{figure}

\subsection{Directed Acyclic Graphs (DAGs)}
A DAG is a graph that comprises variables (called nodes) and arrows in between the nodes (called directed edges) \cite{hassija2020dagiov}. They are arranged in such a way that it is impossible to begin at a node, follow the directed edges in the direction of the arrowhead and land up at the same node at which one started, i.e., it is acyclic in nature \cite{vanderweele2007four}. \textcolor{blue}{It is discussed in \cite{gong2020secured} that DAG-based blockchains such as Nano \cite{lemahieu2018nano}, Byteball \cite{churyumov2016byteball} and IOTA \cite{popov2018tangle} are capable of processing transactions in a parallel manner unlike the conventional sequential way in chain based blockchains. %, as presented in Fig. \ref{fig:normal_vs_DAG}.
This increases the throughput of the blockchain as the chain-based blockchains can process the blocks sequentially one at a time, whereas DAG-based blockchains can process multiple blocks at the same time \cite{yang2020ldv}.} Many studies have proposed security frameworks that use both blockchain and DAG. In \cite{lu2020blockchain}, the authors develop a blockchain architecture that consists of permissioned blockchain and local DAG to enhance the security and reliability of their proposed framework for secure data sharing in the IoV environment. In \cite{hassija2020dagiov}, Hassija et al. used a DAG data structure to store and validate transactions in their proposed framework for V2V communication. Gong et al. \cite{gong2020secured} used a DAG-based approach to improve the security and authenticity of their proposed energy management scheme for smart hybrid micro-grids. Yang et al. used a DAG-based blockchain to ensure high throughput of their proposed framework for Vehicular Social Networks (VSNs) \cite{yang2020ldv}.

\subsection{Cognitive Radio}
A cognitive radio network (CRN) is an environment-aware wireless communication system that operates with two primary goals:
\begin{enumerate}[i.]
    \item Ensuring very high reliable communication wherever and whenever needed. 
    \item Utilizing the available network spectrum efficiently. 
\end{enumerate}

It has been used by researchers to mainly implement efficient spectrum sharing and storage in their frameworks. \textcolor{blue}{Blockchain is used to easily provide security and transparency during the record sharing process within the network.} Rathee et al. use CRN in their proposed framework to provide security to blockchain-enabled IoV networks during spectrum sensing and information transmission \cite{rathee2020crt}. In \cite{nadeem2019securing}, Nadeem et al. use cognitive radio technology to solve the problem of spectrum storage in vehicular systems.

\subsection{Artificial Intelligence (AI)}
AI techniques have been used extensively in many blockchain-based vehicular network schemes for achieving various data and computation-intensive utilities. For example, Deep Neural Networks (DNNs) have been used for evaluation of positioning error in \cite{li2020vehicle} and for accurate vehicle positioning in \cite{song2020blockchain}. Deep Reinforcement Learning (DRL) has been used in many studies for tasks such as \textcolor{black}{modelling} lane changing problem of autonomous vehicles \cite{fu2020autonomous}, node selection \cite{lu2020blockchain}, matching supply and demand during resource sharing \cite{chai2019proof}, and for implementing intelligent and secure content sharing mechanisms \cite{dai2020deep}. Duelling Deep Q-Learning (DDQL) has been used in \cite{zhang2019blockchain} for solving optimization problems and federated learning has been used in \cite{fu2020autonomous} for modelling lane changing problems and in \cite{chai2020hierarchical} for meeting privacy requirements of the IoV network. \textcolor{blue}{In summary, the primary use of AI in blockchain-based systems is to (a) create more intelligent and efficient smart contracts, and (b) create more reliable data that may be used by other members of the blockchain network.}

\subsection{\textcolor{black}{Differential Privacy}}
\textcolor{black}{In differential privacy, some noise is added to the data before query evaluation \cite{hassan2020differential}. The concept of differential privacy has been used in various domains, like IoT data, as it ensures individual privacy, and the addition/deletion of any single record does not affect the final output. The authors of \cite{hassan2019differential} provide a summary of several frameworks that use differential privacy in vehicular networks. When the two concepts of blockchain and differential privacy are integrated, security concerns can be alleviated even in public query evaluation. The authors of \cite{hassan2020differential} provided technical information on integrating differential privacy with blockchain for security concerns, and further show how intelligent transportation systems can use such a framework.}

\subsection{Others}
Few of the frameworks use other emerging technologies for their blockchain-based vehicular networks as follows.

\begin{enumerate}[i.]
    \item UAVs: Su et al. proposed a UAV-assisted aerial to ground vehicular framework for data transmission in disaster-hit areas \cite{su2020lvbs}.
    \item Physical Unclonable Functions (PUFs): PUFs assign a unique ID to an electronic device based on the process variations present in it. This ID can be used for implementing fast and secure authentication, and identification \cite{bansal2020lightweight}. Javaid et al. used PUFs for establishing trust in their proposed blockchain-based framework for vehicular networks \cite{javaid2020scalable}.
    \item Wireless Link Fingerprints: It has been used by Kamal et al. \cite{kamal2020blockchain} in which the channel characteristics of the two vehicles which are communicating with each other generates a unique link. If the variation in the received power of the communicating vehicles is correlating, it means no adversary has affected the communication link. 
    \item 5G: 5G ensures very high-speed communication required by vehicular networks. Gao et al. explore a combination of blockchain and SDN for VANET systems in 5G and fog computing paradigm \cite{gao2019blockchain}.
\end{enumerate}

\subsubsection*{\textcolor{black}{\textbf{Summary}}}
\textcolor{black}{This section discussed a few technologies that have been used recently for securing vehicular networks. The frameworks mentioned here use state-of-the-art technologies like cloud computing, fog computing, edge computing, NDN, SDN, DAG, DNN, DRL, DDQL, cognitive radio, federated learning, differential privacy, 5G, etc. in IoV and VANET networks for security reasons.}

\textcolor{blue}{\section{Major challenges, lessons learnt and future research directions}}
\label{sec:future}
In \textcolor{black}{this} section, we discuss the major challenges faced in implementing blockchain-based applications for securing vehicular networks and the potential future research directions to address them.

\vspace{4mm}
\subsection{Scalability}
Throughput, or the number of transactions validated per second, is a quantitative measure of the scalability of the blockchain system. Bitcoin has a throughput of $7$ transactions per second - for comparison, VISA has a transaction throughput of $2,000$ transactions per second \cite{cromanscaling}. Generally, lower throughputs can be improved at higher scales by suitable modifications to the algorithm itself, such as in \cite{javaid2020scalable}. Vehicular networks will generate a massive amount of data, and because of the latency-critical environment, the blockchain will have to scale up to a very high standard. Many of the existing standards and protocols were developed for cryptocurrency applications, which are not as time-sensitive. \textcolor{blue}{Performance can be increased by choosing the appropriate consensus algorithms and blockchain platforms.} For instance, DPoS and DBFT provide a significant improvement in transaction throughput compared to PoS and PBFT respectively. There has been some research on specifically integrating IoV and blockchain consensus. In \cite{kang2019toward}, the authors posit an enhanced DPoS consensus for blockchain-based IoV applications. Hu et al. \cite{hu2019blockchain} propose an IoV-specific byzantine consensus algorithm for authentication. Consortium and private blockchains are also much more efficient than permissionless blockchains since the number of validating nodes is fewer. \textcolor{blue}{In this case, there must be a careful consideration of the trade-off between decentralization and throughput, since consortium and private blockchains are more centralized than permissionless blockchains.} Another issue related to scalability is the seamless integration of multiple sensors and devices. As the IoV networks become larger, different sensors and platforms will be used. Ensuring they all work together seamlessly will be a challenge.

\subsection{\textcolor{black}{Privacy}}
\textcolor{black}{When the vehicular nodes are used for edge computing, sensitive information like travelling route, card information, etc., are offloaded for various tasks \cite{islam2021blockchain}. To prevent unwanted parties from accessing such information, the information can be encrypted. However, cypher-text makes the analytics process time-consuming. \textcolor{blue}{A speed-up is required in the privacy-preserving blockchain framework for the querying process in edge computing.}}

\textcolor{black}{There is a requirement to ensure user privacy due to the sensitive nature of the data while also allowing transparency to comply with the legal requirements like the General Data Protection Regulation (GDPR) \cite{baldini2020review}. Encryption and complimentary access control techniques are required in a distributed ledger to balance privacy and transparency.}

\subsection{Quantum \textcolor{black}{Computing Attacks}}
Quantum computing is a field of research that will have wide applicability in the coming years \cite{hassija2020forthcoming}. Blockchain relies heavily on the one-way property of its cryptographic hashing techniques. With the advent of quantum computing, these techniques may not be as secure as they are currently. \textcolor{blue}{Quantum computers offer a vastly different scale of computational power; a few quantum computers may easily overcome the computing power of an entire network of ordinary blockchain nodes \cite{fedorov2018quantum}.}
Khalifa et al. \cite{khalifa2019quantum} illustrate how quantum computers may attack proof-of-stake blockchains using both Grover's algorithm and Shor's algorithm, and recommend quantum resilient signature schemes to mitigate them. \textcolor{black}{IoV networks need to be made quantum-resistant, otherwise, they are susceptible to $51\%$ and byzantine attacks \cite{vattaparambil2020scalable}. In an attempt to attack the IoV network using quantum computers, the aim is to affect a part of the network rather than just a single node. This might lead to a loss of trust in the blockchain network.} Quantum attacks is an active area of research, and with many other solutions being proposed \cite{Kiktenko_2018, sun2019towards}, however, these techniques must be incorporated to make commercially viable systems.

\subsection{Prototyping and \textcolor{black}{Simulation}}
Blockchain designed to be a decentralized system poses a natural challenge for prototyping and simulation. \textcolor{blue}{Several large-scale effects cannot be adequately modelled on a prototype.} Existing research works present their findings in the form of small-scale model simulation details, many of which use the best technologies available, however, they do not accurately reflect several unforeseen challenges. For example, it is very difficult to accurately model the randomness that is inherent in vehicular network topologies, even using stochastic mechanisms. \textcolor{blue}{Therefore, the research community would be well aided by the development of better simulation tools and prototype tested frameworks \cite{BC_UAV}.}

\subsection{Attacks on \textcolor{black}{Blockchains}}
The unique nature of a blockchain system opens it up to several attacks that are not of concern in conventional centralized systems - including, but not limited to 51\% attacks, selfish mining, eclipse attacks, DNS attacks, and crypto-jacking \cite{saad2020exploring}. \textcolor{black}{Although blockchain-enabled vehicular edge computing provides benefits like decentralization and transparency, it is vulnerable to several attacks \cite{islam2021blockchain}. In the event of a significant number of members of the network being compromised, the network can be hijacked and the transactions can be forged. The attack occurs due to the lightweight consensus protocols used in permissioned blockchain networks. Hence, a scalable and resilient consensus protocol is required for the deployment of different types of chains in the vehicular edge-computing network.}

\section{Conclusion}
\label{sec:conc}
In this survey, we thoroughly analyzed several blockchain-based security frameworks for vehicular networks and categorised them from three different perspectives, namely, application perspective, security perspective, and blockchain perspective. From the application perspective, we categorise the frameworks based on different application scenarios such as data trading/sharing, resource sharing, parking, traffic management, etc. From the security perspective, we categorise the frameworks based on the security attacks they protect against, the network security requirements they meet, the authentication techniques they employ, and the security proofs they use. From the blockchain perspective, we classify the different frameworks based on the type of blockchain they use, the type of blockchain platform they employ, and the consensus algorithm used in their scheme. We also discussed various simulation tools/platforms which have been used for simulating and testing these blockchain-based frameworks. Furthermore, most of the blockchain-based security frameworks employ other emerging technologies to meet requirements such as low latency, low computation, data storage, etc. in their schemes along with blockchain. Hence it is important to analyze these frameworks and discuss the role of these other state-of-the-art technologies in securing these networks. Lastly, based on the survey, we list out the major challenges and future research directions in this domain. This survey will act as a guide to the researchers and professionals venturing into the research and development of blockchain-based security solutions for vehicular networks like IoV and VANETs.

\bibliography{bibitems.bib}
\bibliographystyle{IEEEtran}

\vskip -1\baselineskip plus -1fil
\begin{IEEEbiography}[{\includegraphics[width=1in,height=1.25in,clip,keepaspectratio]{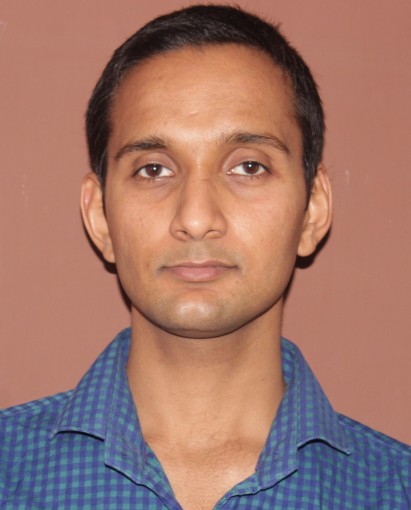}}]{Tejasvi Alladi}\textnormal{ received the B.E. and M.S. degrees from the Birla Institute of Technology and Science, Pilani, India, and North Carolina State University, Raleigh, NC, USA in 2010 and 2015, respectively, and the Ph.D. degree from the Birla Institute of Technology and Science, in 2021. He was a Postdoctoral Researcher with the Department of Systems and Computer Engineering, Carleton University, Ottawa, ON, Canada from Jan 2021-Dec 2021. He is currently an Assistant Professor with the Department of Computer Science and Information Systems, BITS-Pilani. He also has around six years of industrial experience working on embedded systems in Semiconductor MNCs such as Qualcomm Technologies and Samsung Electronics. His research interests include developing security solutions for the Internet of Things (IoT) using cryptography, deep learning, and blockchain technologies.
}
\end{IEEEbiography}

\vskip -1\baselineskip plus -1fil
\begin{IEEEbiography}[{\includegraphics[width=1in,height=1.25in,clip,keepaspectratio]{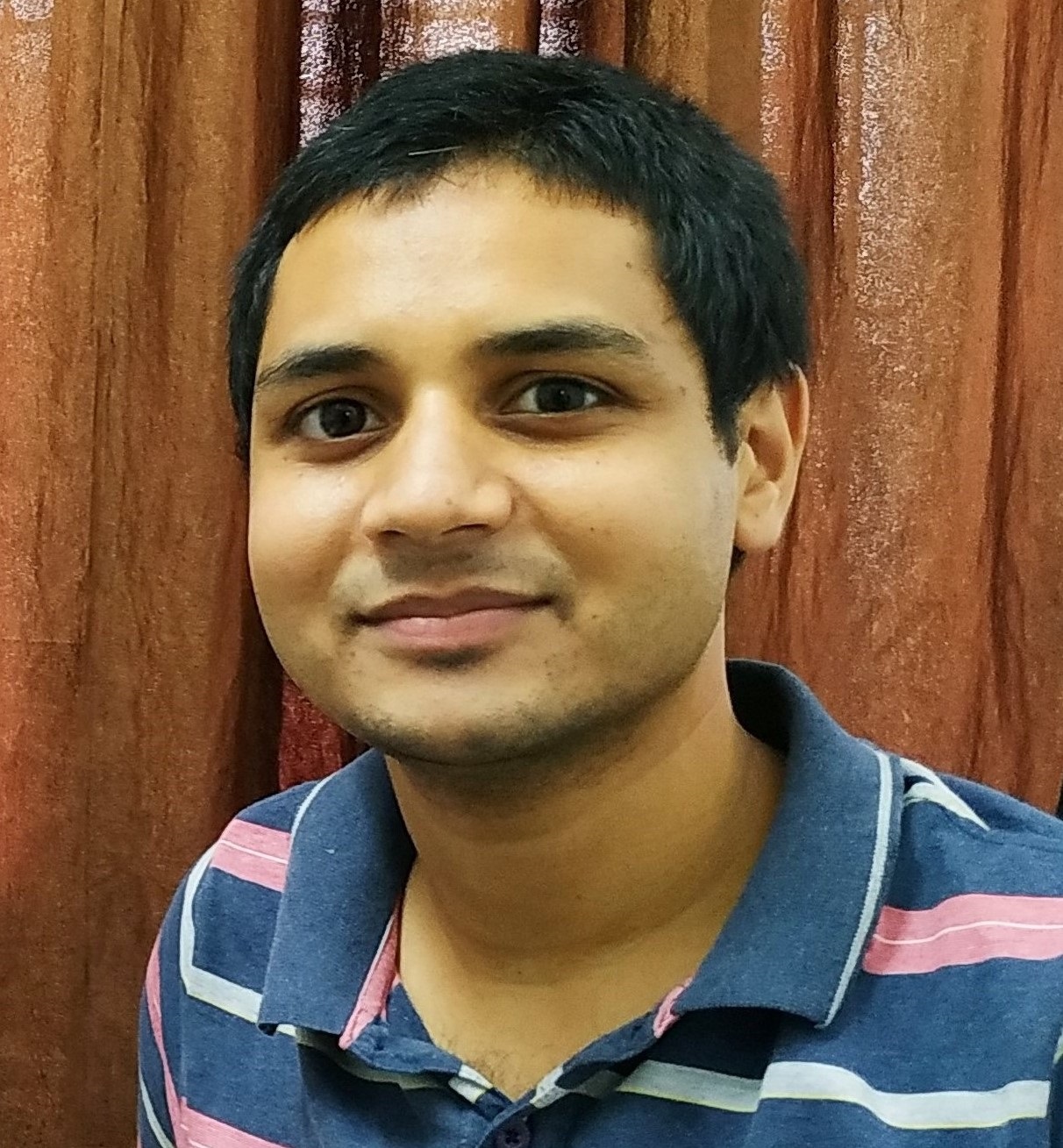}}]{Vinay Chamola}\textnormal{ received the the B.E. degree in electrical and electronics engineering and the master’s degree in communication
engineering from the Birla Institute of Technology
and Science (BITS)-Pilani, Pilani, India, in 2010 and
2013, respectively, and the Ph.D. degree in electrical and computer engineering from the National
University of Singapore, Singapore, in 2016.
In 2015, he was a Visiting Researcher with the
Autonomous Networks Research Group, University
of Southern California at Los Angeles, Los Angeles,
CA, USA. He also worked as a Postdoctoral Research Fellow with the
National University of Singapore. He is currently an Assistant Professor with
the Department of Electrical and Electronics Engineering, BITS-Pilani, where
he heads the Internet of Things Research Group/Lab. He has over 70 publications in high ranked SCI journals, including more than 50 IEEE Transaction,
journal, and magazine articles. His research interests include IoT security,
blockchain, UAVs, VANETs, 5G, and healthcare.
Dr. Chamola serves as an Area Editor for the Ad-Hoc Networks (Elsevier).
He serves as the Co-Chair of various reputed workshops, such as in
IEEE Globecom Workshop 2021, IEEE ANTS 2021, and IEEE ICIAfS
2021. He is a Co-Founder and the President of a Healthcare Startup
Medsupervision Pvt., Ltd. He also serves as an Associate Editor for the IEEE
Internet of Things Magazine, IEEE NETWORKING LETTERS, IET Quantum
Communications, IET Networks, and several other journals. He is a Guest
Editor of Computer Communication (Elsevier), and also the IET Intelligent
Transportation Systems.
}
\end{IEEEbiography}

\vskip -2\baselineskip plus -1fil
\begin{IEEEbiography}[{\includegraphics[width=1in,height=1.5in,clip,keepaspectratio]{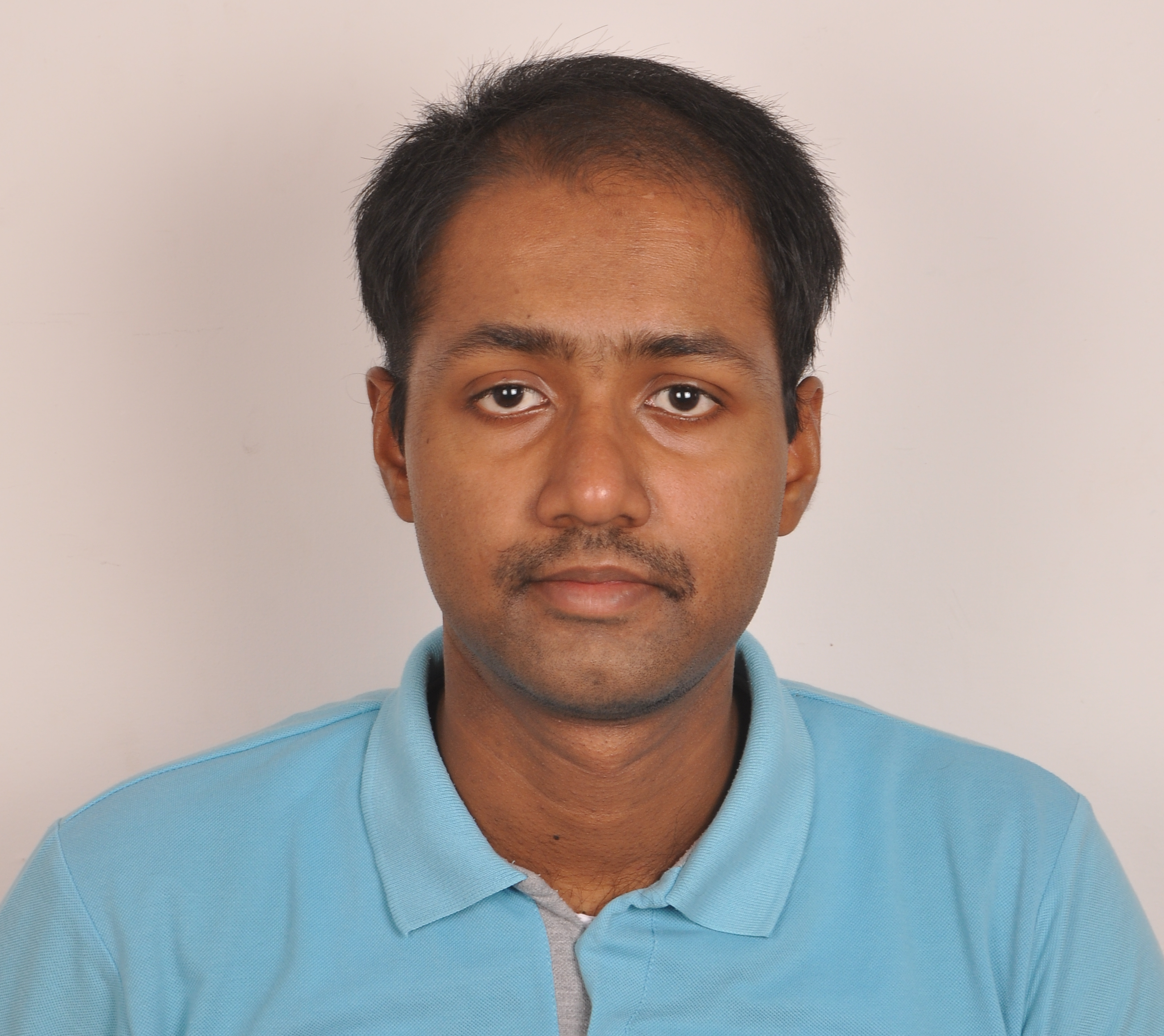}}]{Nishad Sahul} is currently pursuing the M.E. Embedded Systems degree with the Electrical and Electronics Department, Birla Institute of Technology and Science(BITS),Pilani, Pilani Campus. He completed his B.E. in Electrical and Electronics from BITS Pilani, Pilani Campus in 2020. He has co-authored 8 journal papers in SCI-indexed international journals like IEEE Transactions on Industry Applications, IEEE sensors, Vehicular communications etc. and 2 international conference publications. He has interned with two reputed government research laboratories in India; DRDO, Defence Lab, Jodhpur and CSIR-CEERI, Pilani. He has also received many awards like Top Downloaded Paper Certificate 2018-19 by Wiley, Prof. V.S. Rao Best All-rounder award in the 2020 graduating class of BITS Pilani, Pilani Campus, Innovator of the year award 2019-20 by Center for Entrepreneurial Leadership(CEL), BITS Pilani and The Avery Dennison Spirit of Invention scholarship 2017. He received a Student Innovation grant of INR 0.35 million for his master's thesis from AI and Technology Park(ARTPARK), Indian Institute of Science(IISc.) Bangalore. He is also an internationally rated chess player. His research interests include designing hardware accelerators for AI algorithms, vehicular networks, autonomous vehicles, IoT security, Blockchain, FPGAs, UAVs, smart sensor development for IoT applications and system engineering. 
\end{IEEEbiography}

\vskip -2\baselineskip plus -1fil
\begin{IEEEbiography}[{\includegraphics[width=1in,height=1.5in,clip,keepaspectratio]{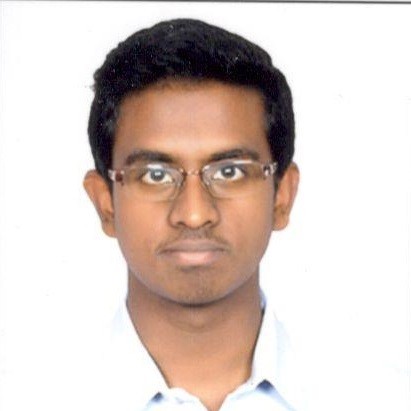}}]{Vishnu Venkatesh} is currently working as an Associate Design Engineer at WCB Robotics Pvt. Ltd. He received his B.E. degree in Electrical and Electronics Engineering from the Birla Institude of Technology and Science, Pilani, in 2021. His primary professional interest is in embedded system development, particularly in hardware and firmware design. He has also worked on research projects and co-authored papers in the fields of cryptography, photonics, IoT security, vehicular networks, and blockchain over the course of his undergraduate degree.
\end{IEEEbiography}

\vskip -2\baselineskip plus -1fil
\begin{IEEEbiography}[{\includegraphics[width=1in,height=1.5in,clip,keepaspectratio]{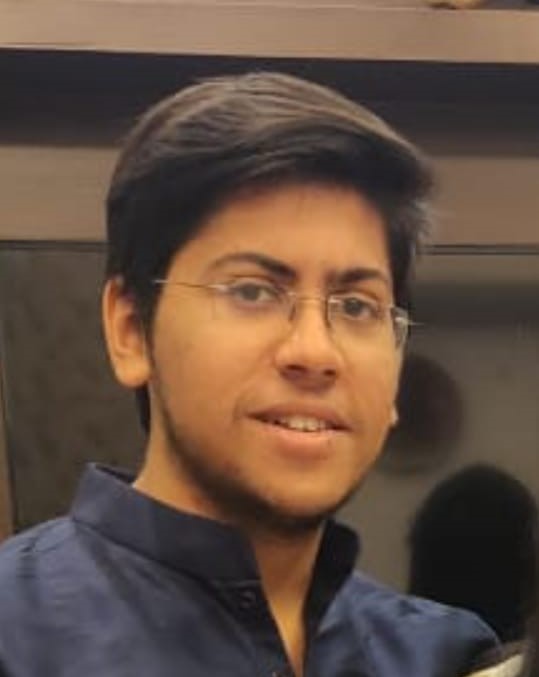}}]{Adit Goyal} is currently pursuing the B.Tech. degree with the Computer Science department, Jaypee Institute of Information Technology (JIIT), Noida. He has completed a few projects in the field of data science, machine learning, and big data. He is currently pursuing his research internship with BITS-Pilani, Pilani, India under Dr. V. Chamola. His research interests include machine learning, data science, and quantum computing.
\end{IEEEbiography}

\vskip -2\baselineskip plus -1fil
\begin{IEEEbiography}[{\includegraphics[width=1in,height=1.25in,clip]{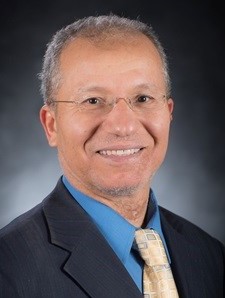}}]{Mohsen Guizani}\textnormal{(S’85–M’89–SM’99–F’09) received the B.S. (with distinction), M.S. and Ph.D. degrees in Electrical and
Computer engineering from Syracuse University, Syracuse, NY,
USA. He is currently a Professor at the Computer Science \&
Engineering Department in Qatar University, Qatar. Previously, he worked in different institutions: University of Idaho, Western Michigan University, University of West Florida, University of Missouri-Kansas City, University of Colorado-Boulder, and Syracuse University. His research interests include wireless communications and mobile computing, applied machine learning, cloud computing, security and its application to healthcare systems. He was elevated to the IEEE Fellow in 2009. He was listed as a Clarivate Analytics Highly Cited Researcher in Computer Science in 2019 and 2020. Dr. Guizani has won several research awards including the “2015 IEEE Communications Society Best Survey Paper Award” as well 4 Best Paper Awards from ICC and Globecom Conferences. He is the author of nine books and more than 800 publications. He is also the recipient of the 2017 IEEE Communications Society Wireless Technical Committee (WTC) Recognition Award, the 2018 AdHoc Technical Committee Recognition Award, and the 2019 IEEE Communications and Information Security Technical Recognition
(CISTC) Award. He served as the Editor-in-Chief of IEEE Network and is currently serves on the Editorial Boards of many IEEE journals/Transactions. He was the Chair of the IEEE Communications Society Wireless Technical Committee and the Chair of the TAOS Technical Committee. He served as the IEEE Computer Society Distinguished Speaker and is currently the IEEE ComSoc Distinguished Lecturer.
}
\end{IEEEbiography}

\end{document}